\documentclass[aps,prl,twocolumn,reprint,showpacs]{revtex4-1}
\usepackage{amsmath}
\usepackage{amssymb}
\usepackage{graphicx}
\usepackage{hyperref}
\usepackage{color,xcolor}
\usepackage{epstopdf}
\begin{document}
\title{Disorder-robust high-field superconducting phase of FeSe single crystals}
\author{Nan Zhou$^1$$^,$$^2$$^,$$^3$}
\author{Yue Sun$^1$$^,$$^4$}
\email{Corresponding author:sunyue@phys.aoyama.ac.jp}
\author{C. Y. Xi$^2$}
\author{Z. S. Wang$^2$}
\author{J. L. Zhang$^2$}
\author{Y. Zhang$^2$}
\author{Y. F. Zhang$^1$}
\author{C. Q. Xu$^1$}
\author{Y. Q. Pan$^1$}
\author{J. J. Feng$^1$}
\author{Y. Meng$^1$}
\author{X. L. Yi$^1$}
\author{L. Pi$^2$}
\author{T. Tamegai$^5$}
\author{Xiangzhuo Xing$^1$}
\email{Corresponding author:xzxing@seu.edu.cn}
\author{Zhixiang Shi$^1$}
\email{Corresponding author:zxshi@seu.edu.cn}

\affiliation{$^1$School of Physics and Key Laboratory of MEMS of the Ministry of Education, Southeast University, Nanjing 211189, China}
\affiliation{$^2$Anhui Province Key Laboratory of Condensed Matter Physics at Extreme Conditions, High Magnetic Field Laboratory, Chinese Academy of Sciences, Hefei 230031, China}
\affiliation{$^3$Institute for Solid State Physics (ISSP), The University of Tokyo, Kashiwa, Chiba 277-8581, Japan}
\affiliation{$^4$Department of Physics and Mathematics, Aoyama Gaguin University, Kanagawa 252-5258, Japan}
\affiliation{$^5$Department of Applied Physics, The University of Tokyo, Tokyo 113-8656, Japan}

\date{\today}

\begin{abstract}

When exposed to high magnetic fields, certain materials manifest an exotic superconducting (SC) phase that has attracted considerable attention. A proposed explanation for the origin of the high-field SC phase is the Fulde-Ferrell-Larkin-Ovchinnikov (FFLO) state. This state is characterized by inhomogeneous superconductivity, where the Cooper pairs have finite center-of-mass momenta. Recently, the high-field SC phase was observed in FeSe, and it was deemed to originate from the FFLO state. Here, we synthesize FeSe single crystals with different levels of disorder. The level of disorder is expressed by the ratio of the mean free path to the coherence length and ranges between 35 and 1.2. The upper critical field \textit{B}$_{\rm{c}2}$ was obtained by both resistivity and magnetic torque measurements over a wide range of temperatures, which went as low as $\sim$0.5 K, and magnetic fields, which went up to $\sim$38 T along the \textit{c} axis and in the \textit{ab} plane. In the high-field region parallel to the \textit{ab} plane, an unusual SC phase was confirmed in all the crystals, and the phase was found to be robust against disorder. This result suggests that the high-field SC phase in FeSe is not a conventional FFLO state.

\end{abstract}

\maketitle

The orbital and Pauli-paramagnetic pair-breaking effects are two distinct mechanisms for destroying superconductivity and limiting the maximum upper critical field in type-\uppercase\expandafter{\romannumeral2} superconductors \cite{werthamer1966temperature, zhou2014controllable}. However, triggered by certain conditions, some unconventional superconductors can easily overcome the Pauli limitation by forming an exotic superconducting (SC) phase \cite{fulde1964superconductivity, larkin1965nonuniform, kitagawa2018evidence, matsuda2007fulde, bianchi2003possible, singleton2000observation, lortz2007calorimetric, bergk2011magnetic, beyer2012angle, koutroulakis2016microscopic, tsuchiya2015phase, agosta2012experimental, wright2011zeeman, mayaffre2014evidence, cho2009upper, coniglio2011superconducting, ok2020observation, kasahara2020evidence, kasahara2014field, cho2017thermodynamic, kenzelmann2008coupled, kenzelmann2010evidence}. Among them, an inhomogeneous SC state occurs when the Pauli pair-breaking effect dominates over the orbital pair-breaking effect, which was independently predicted by Fulde and Ferrell \cite{fulde1964superconductivity} and Larkin and Ovchinnikov \cite{larkin1965nonuniform} (the FFLO state) half a century ago. In the FFLO state, the Zeeman-split Fermi surfaces could drive the formation of Cooper pairs with finite center-of-mass momenta, thus realizing a spatially modulated SC state. The large Maki parameter \textit{$\alpha$}, clean limit (mean free path \textit{$\ell$} $\gg$ coherence length \textit{$\xi$}), and unconventional pairing symmetries \cite{matsuda2007fulde} could drive a system to have easier access to the FFLO state. In previous reports, the FFLO state was suggested to exist in heavy-fermion \cite{kitagawa2018evidence, matsuda2007fulde, bianchi2003possible}, organic \cite{singleton2000observation, lortz2007calorimetric, bergk2011magnetic, beyer2012angle, koutroulakis2016microscopic, tsuchiya2015phase, agosta2012experimental, wright2011zeeman, mayaffre2014evidence, cho2009upper, coniglio2011superconducting}, and some iron-based superconductors \cite{ok2020observation, kasahara2020evidence, kasahara2014field, cho2017thermodynamic}. In the heavy-fermion superconductor CeCoIn$_5$, an additional spin-density wave (SDW) was observed to coexist with SC \cite{kenzelmann2008coupled, kenzelmann2010evidence}, implying that the high-field SC phase does not simply originate from the FFLO state. In contrast, in some layered organic superconductors, the FFLO state was observed without any magnetic order \cite{singleton2000observation, lortz2007calorimetric, bergk2011magnetic, beyer2012angle, koutroulakis2016microscopic, tsuchiya2015phase, agosta2012experimental, wright2011zeeman, mayaffre2014evidence, cho2009upper, coniglio2011superconducting}.

\begin{figure*}
\includegraphics[width=42.5pc]{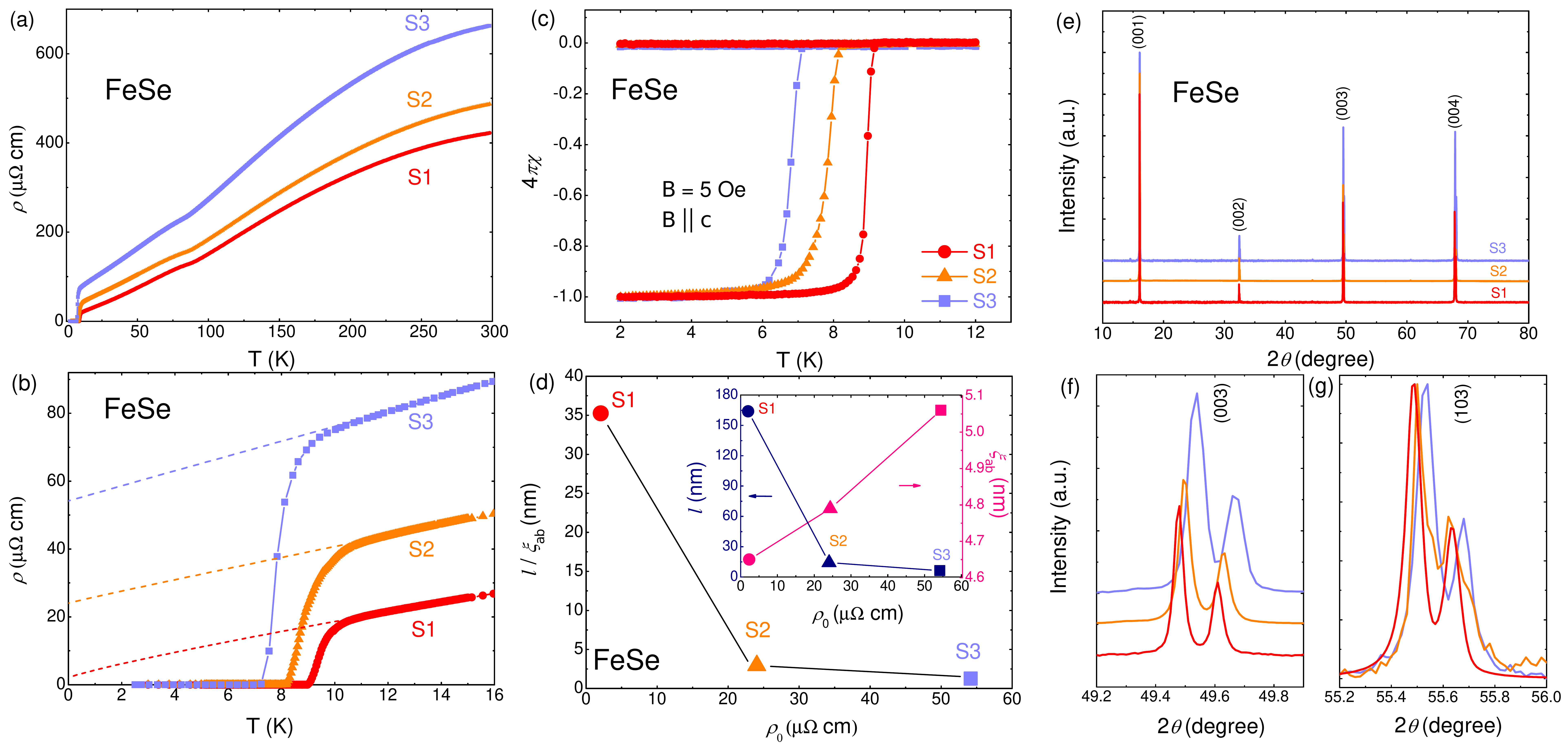}
\begin{center}
\caption{\label{fig1} Temperature dependences of the resistivity at zero field at temperatures (a) in the range 0-300 K and (b) below 16 K for three selected FeSe single crystals with different amounts of disorder. The dashed lines correspond to the power-law fitting $\rho$(\textit{T}) = $\rho$$_0$+\textit{A}\textit{T}$^{\textit{$\alpha$}}$. (c) Temperature dependence of magnetic susceptibility \textit{$\chi$} measured under 5 Oe field along the \textit{c} axis. (d) Residual resistivity \textit{$\rho$}$_{0}$ dependence of the ratio of the mean free path \textit{$\ell$} to the coherence length \textit{$\xi$}$_{\textit{ab}}$. The inset displays the \textit{$\rho$}$_{0}$ dependence of \textit{$\ell$} (left axis) and the \textit{$\xi$}$_{\textit{ab}}$ (right axis). (e) The single-crystal XRD patterns for the three selected FeSe single crystals. Enlarged diffraction peaks of (f) (003) and (g) (103).}
\end{center}
\end{figure*}

The target material of this study is the iron chalcogenide superconductor FeSe \cite{hsu2008superconductivity}, which has attracted considerable interest due to its unique electrical properties, such as a multiband structure \cite{watson2015dichotomy}, electronic nematicity \cite{hosoi2016nematic}, extremely small Fermi energy \cite{kasahara2014field}, and a strongly orbital-dependent pairing mechanism \cite{sprau2017discovery}. Recently, a high-field SC phase was observed in FeSe at temperatures below 2 K and under applied field close to the upper critical field \cite{kasahara2020evidence, ok2020observation}. It has been suggested that the high-field SC phase may originate from the FFLO state \cite{kasahara2020evidence, ok2020observation}.

Since the phase transition from FFLO to the BCS state is first order, the FFLO state is sensitive to disorder \cite{aslamazov1969influence, takada1970superconductivity, matsuda2007fulde}. Therefore, the disorder effect can be a useful method to clarify whether the high-field SC phase in FeSe stems from the FFLO state. In this study, we probed the high-field SC phase in three FeSe single crystals with a disorder level \textit{$\ell$}/\textit{$\xi$} ranging from 35 to 1.2. We found that the high-field SC phase exists in all three crystals and is robust against disorder.

The single crystals of FeSe studied here were synthesized by the vapor transport method \cite{sun2016electron,bohmer2016variation}. Single crystals with different amounts of disorder were selected from different batches. The structure of the FeSe single crystals was characterized by x-ray diffraction (XRD) with Cu \textit{K}\textit{$\alpha$} radiation. The crystal composition was determined by energy-dispersive x-ray spectroscopy (EDX). The temperature dependence of the resistivity up to 9 T was measured using a physical property measurement system (PPMS, Quantum Design). The magnetization was measured by a commercial SC quantum interference device magnetometer (MPMS-XL5, Quantum Design). The high-field transport and magnetic torque measurements were carried out in a water-cooled magnet with steady fields up to 38 T at the High Magnetic Field Laboratory of the Chinese Academy of Sciences by using standard a.c. lock-in and capacitive cantilever techniques, respectively.

Figure 1(a) shows the zero-field resistivity \textit{$\rho$}(\textit{T}) for the three selected FeSe single crystals. As shown in Fig. 1(b), the SC transition temperatures $T_{\rm{c}}$ determined by the zero resistivity are $\sim$9.0 K (sample S1), 8.2 K (sample S2), and 7.1 K (sample S3). The residual resistivity \textit{$\rho$}$_{0}$ is determined by using the power-law fitting $\rho$(\textit{T}) = $\rho$$_0$+\textit{A}\textit{T}$^{\textit{$\alpha$}}$ (\textit{$\rho$}$_{0}$, \textit{A}, and \textit{$\alpha$} are the fitting parameters) from normal state data to zero temperature as shown by the dashed lines in Fig. 1(b). The obtained \textit{$\rho$}$_{0}$ are $\sim$2.1 $\mu\Omega$ cm (sample S1), 24 $\mu\Omega$ cm (sample S2) and 54.2 $\mu\Omega$ cm (sample S3). The residual resistivity ratio (RRR), defined as \textit{$\rho$}$_{300 K}$/\textit{$\rho$}$_{0}$, is estimated to be $\sim$207 for sample S1, $\sim$20.3 for sample S2, and $\sim$12.2 for sample S3 (Table I). A kink like behavior at \textit{$T$} $\sim$ 90 K related to the structural transition \cite{mcqueen2009tetragonal} can be seen more clearly in the plot of \textit{$d$}\textit{$\rho$}/\textit{$d$}\textit{T} (see Fig. S1 \cite{supplement}). The structural transition temperature $T_{\rm{s}}$ is found to be slightly suppressed by the disorder, which is consistent with previous reports \cite{bohmer2016variation}.

\begin{table*}
  \centering
  \caption{Summary of the experimentally derived parameters for samples S1, S2, and S3. $T_{\rm{c}}$, defined by the onset of resistivity; \textit{$\rho$}$_{0}$, residual resistivity; RRR, residual resistivity ratio; \textit{$\alpha$}, Maki parameter; \textit{$B$}$_{\rm{c}2}^{\textit{c}}$(0) and \textit{$B$}$_{\rm{c}2}^{\textit{ab}}$(0), the estimated upper critical fields along the \textit{c} axis and in the \textit{ab} plane; \textit{$\ell$}, the mean free path; \textit{$\xi$}$_{\textit{c}}$(0) and \textit{$\xi$}$_{\textit{ab}}$(0), the coherence length along the \textit{c} axis and in the \textit{ab} plane, determined from \textit{$B$}$_{\rm{c}2}^{\textit{c}}$(0) and \textit{$B$}$_{\rm{c}2}^{\textit{ab}}$(0); and \textit{$\ell$}/\textit{$\xi$}$_{\textit{ab}}$(0), the ratio of the mean free path \textit{$\ell$} to the coherence length \textit{$\xi$}$_{\textit{ab}}$(0).}
  \label{oscillations}
  \begin{tabular*}{1\textwidth}{@{\extracolsep{\fill}}cccccccccccc}
    \hline
    \hline
     & $T_{\rm{c}}$(K) & \textit{$\rho$}$_{0}$($\mu\Omega$ cm) & RRR & \textit{$\alpha$} & \textit{$B$}$_{\rm{c}2}^{\textit{c}}$(0)(T) & \textit{$B$}$_{\rm{c}2}^{\textit{ab}}$(0)(T) & \textit{$\ell$} (nm) & \textit{$\xi$}$_{\textit{c}}$(0)(nm) & \textit{$\xi$}$_{\textit{ab}}$(0)(nm) & \textit{$\ell$}/\textit{$\xi$}$_{\textit{ab}}$(0) \\
    \hline
    S1  & 9.0 & 2.1 & 207 & 1.13 & 15.26 & 27.21 & 164 & 2.6 & 4.7 & 35.2\\
    S2  & 8.2 & 24 & 20.3 & 1.32 & 14.33 & 25.28 & 14.1 & 2.7 & 4.8 & 3.0\\
    S3  & 7.1 & 54.2 & 12.2 & 1.48 & 12.86 & 22.27 & 6.3 & 2.9 & 5.1 & 1.2\\
    \hline
    \hline
  \end{tabular*}
\end{table*}

Superconductivity was confirmed by the temperature dependence of the susceptibility measurements, as shown in Fig. 1(c). The obtained $T_{\rm{c}}$, which is defined as the deviation of the zero-field-cooling and field-cooling susceptibilities, is in good agreement with the resistivity data. The sharp SC transition width indicates the homogeneous distribution of disorder.

We estimated the mean free path by assuming that the hole and electron pockets are compensated perfectly. According to the expression \textit{$\ell$} = $\frac{\pi{c}\hbar}{{N}{e^2}{k_F}{\rho_0}}$ \cite{kasahara2020evidence}, where \textit{c} is the lattice parameter, \textit{N} is the number of formula units per unit cell, \textit{$k$}$_{\rm{F}}$ = 1.07 nm$^{-1}$ \cite{kasahara2020evidence} is the Fermi wave vector, and \textit{$\rho$}$_{0}$ is the residual resistivity, \textit{$\ell$} is estimated to be $\sim$164, 14.1, and 6.3 nm for samples S1, S2, and S3, respectively [see the inset in Fig. 1(d)]. The ratio of the mean free path \textit{$\ell$} to the in-plane coherence length \textit{$\xi$}$_{\textit{ab}}$ is presented in the main panel of Fig. 1(d) and Table I. The coherence length is estimated from \textit{B}$_{\rm{c}2}$, as discussed in the Supplemental Material (see Figs. S4 and S5 \cite{supplement}). Our results confirm that sample S1 is in the clean limit with \textit{$\ell$}/\textit{$\xi$}$_{\textit{ab}}$ $>$ 35. By contrast, considerable amounts of disorder have been successfully introduced into samples S2 and S3 because \textit{$\ell$}/\textit{$\xi$}$_{\textit{ab}}$ is reduced to $\sim$3.0 and 1.2, respectively.

\begin{figure*}
\includegraphics[width=42.5pc]{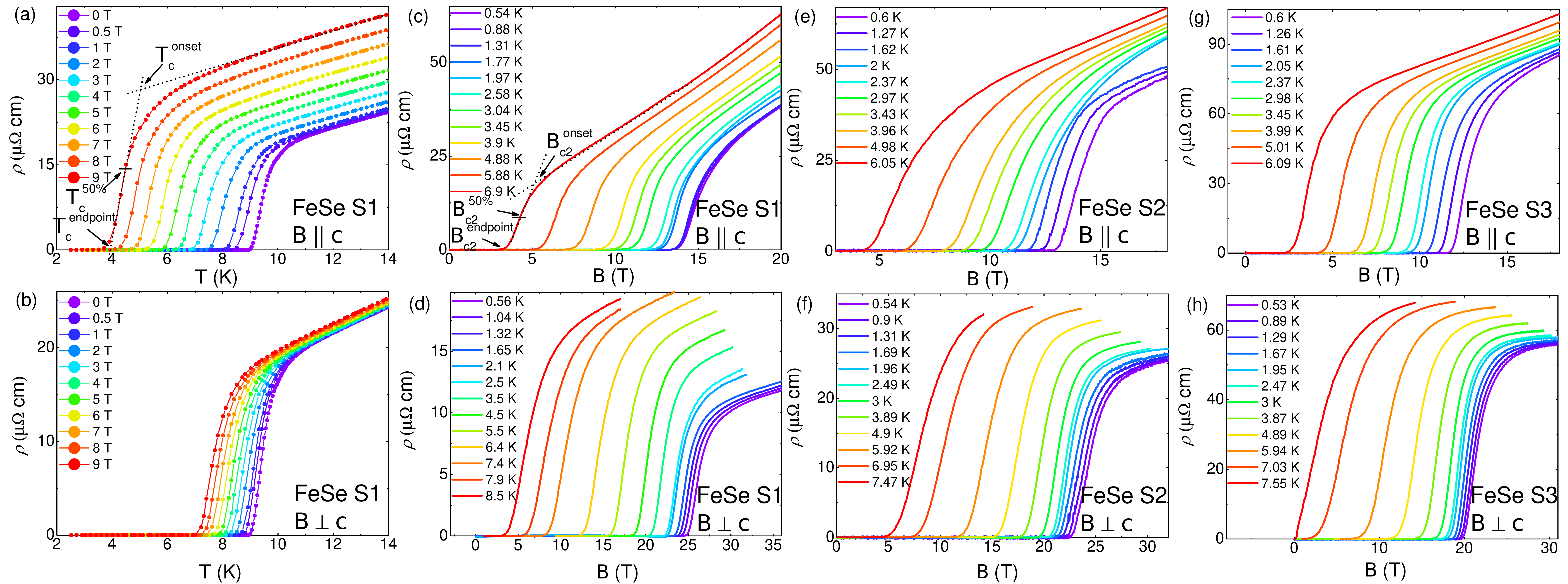}
\begin{center}
\caption{\label{fig2} Temperature dependence of the resistivity for sample 1 with (a) \textit{B} $\parallel$ \textit{c} and (b) \textit{B} $\perp$ \textit{c} under fields from 0 to 9 T. Magnetic field dependence of the resistivity with \textit{B} $\parallel$ \textit{c} and \textit{B} $\perp$ \textit{c} at several temperatures for (c) and (d) sample S1, (e) and (f) sample S2, and (g) and (h) sample S3.}
\end{center}
\end{figure*}

We conducted structure and composition analyses to obtain further information about the disorder in the three crystals [see Fig. 1(e)]. Only the (00\textit{$\ell$}) peaks are observed for all three crystals, which can be well indexed based on a tetragonal structure with the \textit{P}4/\textit{nmm} space group. The positions of peaks were found to be slightly shifted to a higher angle from S1 to S3, which can be seen more clearly in the enlarged view of (003) peaks shown in Fig. 1(f). To obtain the lattice constant \textit{$a$}/\textit{$b$}, we measured (103) peaks by scanning the crystal angle \textit{$\omega$} independently of 2\textit{$\theta$} (the angle between incident and scattered x-rays), as shown in Fig. 1(g). The lattice constants are estimated to be \textit{$a$} = (3.777$\pm$0.002) {\AA}, \textit{$c$} = (5.524$\pm$0.002) {\AA} for sample S1; \textit{$a$} = (3.765$\pm$0.002) {\AA}, \textit{$c$} = (5.521$\pm$0.002) {\AA} for sample S2; and \textit{$a$} = (3.761$\pm$0.002) {\AA}, \textit{$c$} = (5.519$\pm$0.002) {\AA} for sample S3. Both \textit{$a$} and \textit{$c$} decrease with increasing disorder, indicating lattice shrinkage. The EDX measurements show that the molar ratio of Fe:Se is $\sim$1:1.005, 1:1.074, and 1:1.087 for samples S1, S2, and S3, respectively, which indicates that the amount of Fe is less than the amount of Se and the proportion of Fe decreases with increasing disorder. The lattice shrinkage and the smaller amount of Fe suggest that the disorder could be Fe vacancies, which were confirmed by scanning tunneling microscopy (STM) measurements \cite{jiao2017impurity}. Furthermore, the Fe vacancies were demonstrated to be non-magnetic disorder \cite{jiao2017impurity}.

\begin{figure}
\includegraphics[width=20pc]{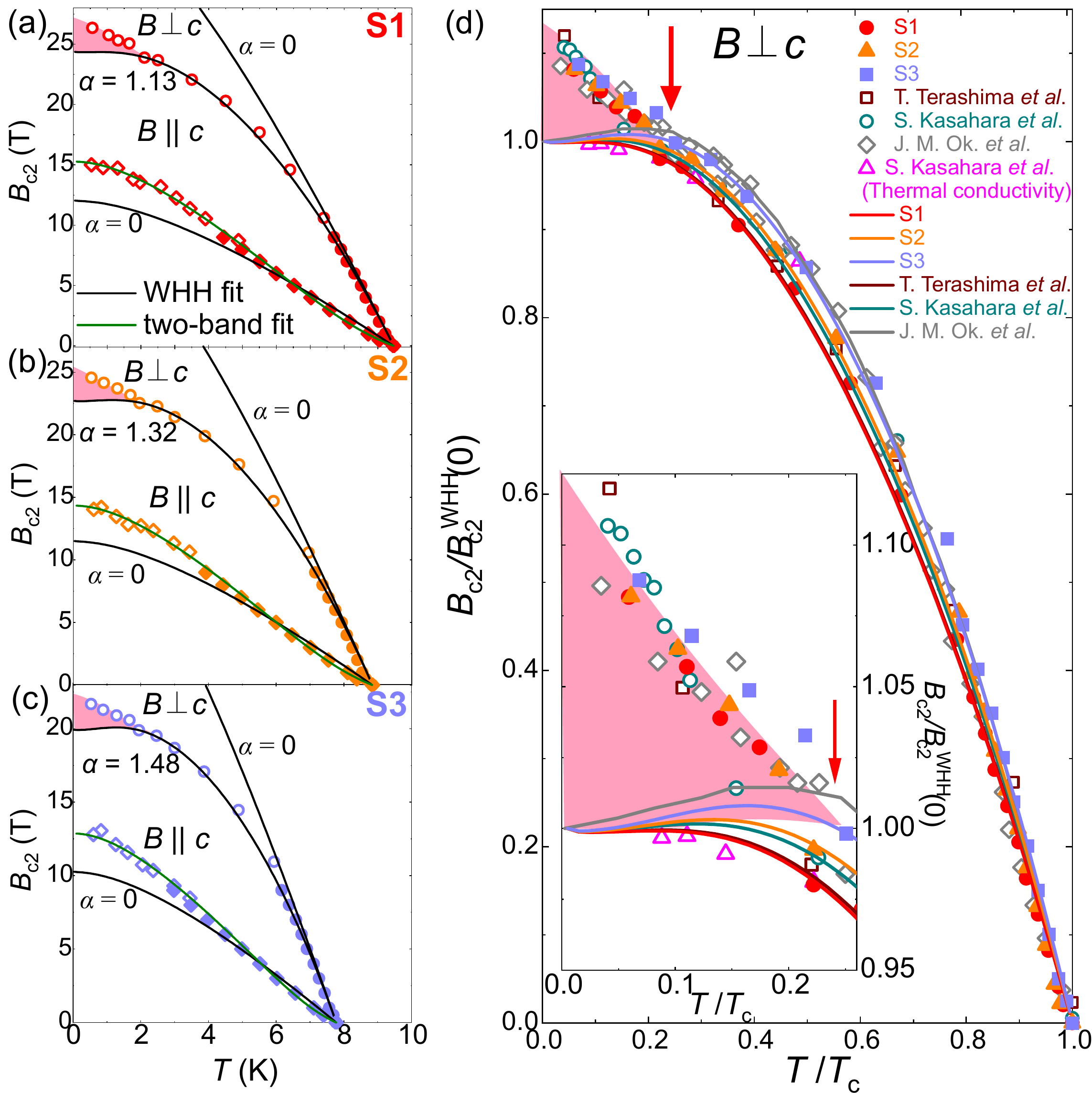}
\begin{center}
\caption{\label{fig3} [(a)--(c)] Temperature dependences of the upper critical field \textit{B}$_{\rm{c}2}$(\textit{T}) for the three selected FeSe single crystals. Diamonds and circles represent the cases with \textit{B} $\parallel$ \textit{c} and \textit{B} $\perp$ \textit{c}, respectively. The two-band model (green lines) and WHH fitting curves (black lines) are shown for \textit{B} $\parallel$ \textit{c} and \textit{B} $\perp$ \textit{c}. The WHH model predictions with \textit{$\alpha$} = 0 are also presented for comparison. (d)
The upper critical fields \textit{$B$}$_{\rm{c}2}^{\textit{ab}}$ obtained with the resistivity reported by Terashima \textit{et al}.  \cite{terashima2014anomalous} (open squares), Kasahara \textit{et al}. \cite{kasahara2020evidence} (open circles), and Ok \textit{et al}. \cite{ok2020observation} (open diamonds); obtained with the thermal conductivity reported by Kasahara \textit{et al}. (open triangles) \cite{kasahara2020evidence}; and obtained from the crystals used in this paper are normalized by \textit{$B$}$_{\rm{c}2}^{\rm{WHH}}$(0 K) for comparison. The solid lines represent the WHH fitting. The inset shows an enlarged view of the low-temperature part. The critical fields obtained with resistivity represent the upper boundary between the high-field SC phase and the normal state, while the critical fields obtained with thermal conductivity and the WHH fitting represent the lower boundary between the high-field SC phase and the BCS state. The pink shaded area represents the region for the high-field SC phase.}
\end{center}
\end{figure}

We measured the temperature dependence of resistivity under fields ranging from 0 to 9 T and the field dependence of resistivity up to 38 T at several temperatures to probe the upper critical fields of the three crystals (see Figs. 2 and S2 \cite{supplement}). The determinations of \textit{B}$_{\rm{c}2}$ by the criteria of onset, 50\%, and the end point of the SC transition are shown in Figs. 2(a) and 2(c). \textit{B}$_{\rm{c}2}$ obtained with different criteria show similar behavior. In the following, \textit{B}$_{\rm{c}2}$ determined with the criterion of 50\% of the SC transition is adopted for discussion since it is less affected by the SC fluctuation and the vortex motion. However, \textit{B}$_{\rm{c}2}$ obtained with the criteria of the onset and end point of the SC transition are presented in Figs. S3 and S8 \cite{supplement}. Here, we want to emphasize that our conclusion is not affected by the criteria of \textit{B}$_{\rm{c}2}$, which will be shown below. To further confirm the reliability of the obtained \textit{B}$_{\rm{c}2}$, we also performed the thermodynamic measurements of magnetic torque \cite{li2007low, kasahara2012electronic}.
The irreversibility field \textit{B}$_{\rm{irr}}$, defined as the onset of the separating field for the up- and down-sweep torque data (as indicated by arrows in Figs. S7(a)-S7(l) \cite{supplement}), is in good agreement with the upper critical field \textit{B}$_{\rm{c}2}^{\rm{endpoint}}$ (see Figs. S8(a)--S8(c) \cite{supplement}).


For \textit{B} $\parallel$ \textit{c}, the experimental data \textit{B}$_{\rm{c}2}^{\textit{c}}$ for the three crystals are well above the predicted upper limit based on the Werthamer-Helfand-Hohenberg (WHH) theory \cite{werthamer1966temperature} with Maki parameter \textit{$\alpha$} = 0 and spin orbit interaction \textit{$\lambda$} = 0 (shown by the black lines in Fig. 3) and manifest a concave increase at \textit{$T$} $>$ 4 K but a convex increase at \textit{$T$} $<$ 4 K. Similar behavior has also been observed in other iron-based superconductors \cite{wang2020strong, xing2017two}, which can be well reproduced by the two-band model (shown by the green lines). For \textit{$B$} $\parallel$ \textit{ab}, the experimental data \textit{$B$}$_{\rm{c}2}^{\textit{ab}}$ fall below the WHH prediction at low temperatures with \textit{$\alpha$} = 0 and \textit{$\lambda$}$_{\rm{so}}$ = 0, indicating that the spin-paramagnetic effect cannot be ignored. By considering a finite \textit{$\alpha$} (\textit{$\lambda$} is kept at 0), the WHH model can fit \textit{$B$}$_{\rm{c}2}^{\textit{ab}}$ well in the temperature range of \textit{$T$} $>$ 2 K. The Maki parameter \textit{$\alpha$} extracted from the WHH fitting is plotted in Fig. S5(e) \cite{supplement}, which is increased with \textit{$\rho$}$_{0}$ (Table I). The anisotropy \textit{$\gamma$} = \textit{$B$}$_{\rm{c}2}^{\textit{ab}}$/\textit{$B$}$_{\rm{c}2}^{\textit{c}}$ is found to be slightly increased in crystals with more disorder (see Fig. S5(d) \cite{supplement}).

Interestingly, at \textit{$T$} $<$ 2 K, the experimental data significantly deviate from the WHH curves and display an unusual upturn for all three crystals. This phenomenon was also observed in previous studies and was predicted to originate from a new high-field SC phase \cite{ok2020observation, kasahara2020evidence}. To directly compare the high-field SC phase for crystals with different amounts of disorder, \textit{$B$}$_{\rm{c}2}^{\textit{ab}}$ normalized by \textit{$B$}$_{\rm{c}2,\textit{ab}}^{\rm{WHH}}$(0 K) (WHH fitting with finite \textit{$\alpha$}) are shown in Fig. 3(d) (criterion of 50\%) and Fig. S8(d) (criterion of the end point) \cite{supplement}. The upper critical fields \textit{$B$}$_{\rm{c}2}^{\textit{ab}}$ obtained with the resistivity reported by Terashima \textit{et al}. (open squares) \cite{terashima2014anomalous}, Kasahara \textit{et al}. (open circles) \cite{kasahara2020evidence}, and Ok \textit{et al}. (open diamonds) \cite{ok2020observation}; obtained with the thermal conductivity reported by Kasahara \textit{et al}. (open triangles) \cite{kasahara2020evidence}; and obtained with the heat capacity reported by Hardy \textit{et al}. \cite{hardy2020vortex} [open pentagons, only shown in Fig. S8(d) since the vortex melting temperature corresponding to the \textit{$B$}$_{\rm{c}2}^{\rm{endpoint}}$] are also incorporated for comparison. According to the previous reports, the critical fields obtained with resistivity and torque measurements represent the upper boundary between the high-field SC phase and the normal state, while the critical fields obtained with thermal conductivity represent the lower boundary between the high-field SC phase and the BCS state. The lower boundary estimated with thermal conductivity is roughly consistent with that obtained from the WHH fitting. Crystals used in different reports are supposed to contain different amounts of disorder. Obviously, the upturn behavior was observed in \textit{$B$}$_{\rm{c}2}^{\textit{ab}}$ for all the crystals. Furthermore, the enhancements of \textit{$B$}$_{\rm{c}2}^{\textit{ab}}$ above the lower boundary, i.e., the region for the high-field SC phase (marked by the pink shading), and the tricritical point of the normal, BCS, and high-field SC phase (indicated by the arrow) for different crystals are almost identical, which can be seen more clearly in the inset of Fig. 3(d) and in Fig. S8(d) \cite{supplement}.


For a conventional \textit{s}-wave superconductor, the FFLO state is well known to be very sensitive to nonmagnetic disorder, which exists only in crystals in the clean limit \cite{aslamazov1969influence, takada1970superconductivity, matsuda2007fulde}. On the other hand, it has been proposed using the theoretical calculation that the FFLO state could be less susceptible to disorder in an unconventional superconductor, such as the disordered \textit{s}-wave superconductors \cite{cui2008fulde} or \textit{d}-wave superconductors \cite{vorontsov2005phase, ptok2010fulde}. In this case, the order parameter is changed in the FFLO state, and the phase diagram is quite different from the conventional one \cite{agterberg2001effect, wang2007impurity}. FeSe is well accepted to have the sign-reversed \textit{s}$_{\pm}$ pairing, with nodes or gap minima in multigaps \cite{sprau2017discovery}. The unique pairing mechanism may also make FeSe less susceptible to disorder, similar to the case of \textit{d}-wave superconductors. Therefore, although the high-field SC state in FeSe is found to be robust to disorder, we cannot simply exclude the possibility of the FFLO state. Theoretical study of the disorder effect on the FFLO state in a superconductor with \textit{s}$_{\pm}$ pairing is required to solve this issue, although the phase transition from the BCS to FFLO state in \textit{s}$_{\pm}$ pairing has been theoretically predicted to be first order \cite{ptok2014influence}.

Another possible origin of the high-field SC phase is the coexistence of the SDW order, which is triggered by the nesting effect around the nodal position of the SC gap \cite{kenzelmann2017exotic}. The SDW order has been discussed as another possible mechanism for the high-field SC phase in the heavy-fermion superconductor CeCoIn$_5$ besides the FFLO state \cite{kenzelmann2008coupled, kenzelmann2017exotic, suzuki2011theory}. A theoretical calculation proposed that the SDW order close to \textit{$B$}$_{\rm{c}2}$ is a direct result of the strong spin-paramagnetic pair-breaking effect and nodal gap structure \cite{ikeda2010antiferromagnetic}. In FeSe, the gap nodes or deep minima in both the electron-type \textit{$\epsilon$} band with a small gap size and the hole-type \textit{$\alpha$} band with a large gap size have already been confirmed \cite{sprau2017discovery, kasahara2014field, sun2017gap}, which makes the field-induced SDW order possible. Such SDW order coexisting with superconductivity should be sensitive to the gap structure. If the nodes or gap minima are smeared out, such a SDW order will disappear spontaneously. Our previous work observed that the nodes or gap minima in the small \textit{$\epsilon$} gap can be smeared out by disorder \cite{sun2018disorder}. However, the nodes or gap minima in the large \textit{$\alpha$} gap should be more robust against disorder because the V-shaped spectrum was observed in the STM measurements in the FeSe crystals with a high level of S doping \cite{hanaguri2018two}. Since the high-field SC phase in FeSe is observed only at fields close to \textit{$B$}$_{\rm{c}2}$, it should be attributed to the larger \textit{$\alpha$} gap. Therefore, our observation of the disorder-robust behavior is not contradictory to the SDW order induced in the high-field SC phase. Further efforts, such as nuclear magnetic resonance and neutron diffraction measurements, are required to clarify the origin of the high-field SC phase in FeSe. Nevertheless, our observations suggest that FeSe provides an intriguing platform to study the interplay between multiple phases such as superconductivity, nematicity, and the SDW or FFLO state.

To conclude, we studied the upper critical fields of FeSe single crystals with different amounts of disorder. A high-field SC phase was observed in all the crystals, and it was found to be robust against disorder. These results suggest that the high-field SC state in FeSe is related to a disorder-robust order, which may provide new clues to understand the exotic properties of FeSe.

The authors would like to thank S. Kasahara, A. Ptok, R. Ikeda and C. Meingast for their helpful discussions. A portion of this work was performed at the Steady High Magnetic Field Facilities, High Magnetic Field Laboratory, Chinese Academy of Sciences, and supported by the High Magnetic Field Laboratory of Anhui Province. This work was partly supported by the National Key R$\&$D Program of China (Grant No. 2018YFA0704300), the Strategic Priority Research Program (B) of the Chinese Academy of Sciences (Grant No. XDB25000000), the National Natural Science Foundation of China (Grants No. U1932217, No. 11674054, and No. 11874359), the China Postdoctoral Science Foundation (Grant No. 2019M661679), and JSPS KAKENHI (Grants No. JP20H05164, No. JP19K14661, and No. JP17H01141).

N.Z. and Y.S. contributed equally to this paper.

\bibliographystyle{apsrev4-1}
\bibliography{FeSeFFLO}

\begin{thebibliography}{53}%
\makeatletter
\providecommand \@ifxundefined [1]{%
 \@ifx{#1\undefined}
}%
\providecommand \@ifnum [1]{%
 \ifnum #1\expandafter \@firstoftwo
 \else \expandafter \@secondoftwo
 \fi
}%
\providecommand \@ifx [1]{%
 \ifx #1\expandafter \@firstoftwo
 \else \expandafter \@secondoftwo
 \fi
}%
\providecommand \natexlab [1]{#1}%
\providecommand \enquote  [1]{``#1''}%
\providecommand \bibnamefont  [1]{#1}%
\providecommand \bibfnamefont [1]{#1}%
\providecommand \citenamefont [1]{#1}%
\providecommand \href@noop [0]{\@secondoftwo}%
\providecommand \href [0]{\begingroup \@sanitize@url \@href}%
\providecommand \@href[1]{\@@startlink{#1}\@@href}%
\providecommand \@@href[1]{\endgroup#1\@@endlink}%
\providecommand \@sanitize@url [0]{\catcode `\\12\catcode `\$12\catcode
  `\&12\catcode `\#12\catcode `\^12\catcode `\_12\catcode `\%12\relax}%
\providecommand \@@startlink[1]{}%
\providecommand \@@endlink[0]{}%
\providecommand \url  [0]{\begingroup\@sanitize@url \@url }%
\providecommand \@url [1]{\endgroup\@href {#1}{\urlprefix }}%
\providecommand \urlprefix  [0]{URL }%
\providecommand \Eprint [0]{\href }%
\providecommand \doibase [0]{http://dx.doi.org/}%
\providecommand \selectlanguage [0]{\@gobble}%
\providecommand \bibinfo  [0]{\@secondoftwo}%
\providecommand \bibfield  [0]{\@secondoftwo}%
\providecommand \translation [1]{[#1]}%
\providecommand \BibitemOpen [0]{}%
\providecommand \bibitemStop [0]{}%
\providecommand \bibitemNoStop [0]{.\EOS\space}%
\providecommand \EOS [0]{\spacefactor3000\relax}%
\providecommand \BibitemShut  [1]{\csname bibitem#1\endcsname}%
\let\auto@bib@innerbib\@empty
\bibitem [{\citenamefont {Werthamer}\ \emph {et~al.}(1966)\citenamefont
  {Werthamer}, \citenamefont {Helfand},\ and\ \citenamefont
  {Hohenberg}}]{werthamer1966temperature}%
  \BibitemOpen
  \bibfield  {author} {\bibinfo {author} {\bibfnamefont {N.}~\bibnamefont
  {Werthamer}}, \bibinfo {author} {\bibfnamefont {E.}~\bibnamefont {Helfand}},
  \ and\ \bibinfo {author} {\bibfnamefont {P.}~\bibnamefont {Hohenberg}},\
  }\href@noop {} {\bibfield  {journal} {\bibinfo  {journal} {Phys. Rev.}\
  }\textbf {\bibinfo {volume} {147}},\ \bibinfo {pages} {295} (\bibinfo {year}
  {1966})}\BibitemShut {NoStop}%
\bibitem [{\citenamefont {Zhou}\ \emph {et~al.}(2014)\citenamefont {Zhou},
  \citenamefont {Xu}, \citenamefont {Wang}, \citenamefont {Yang}, \citenamefont
  {Li}, \citenamefont {Guo}, \citenamefont {Yang}, \citenamefont {Niu},
  \citenamefont {Chen}, \citenamefont {Cao} \emph
  {et~al.}}]{zhou2014controllable}%
  \BibitemOpen
  \bibfield  {author} {\bibinfo {author} {\bibfnamefont {N.}~\bibnamefont
  {Zhou}}, \bibinfo {author} {\bibfnamefont {X.}~\bibnamefont {Xu}}, \bibinfo
  {author} {\bibfnamefont {J.}~\bibnamefont {Wang}}, \bibinfo {author}
  {\bibfnamefont {J.}~\bibnamefont {Yang}}, \bibinfo {author} {\bibfnamefont
  {Y.}~\bibnamefont {Li}}, \bibinfo {author} {\bibfnamefont {Y.}~\bibnamefont
  {Guo}}, \bibinfo {author} {\bibfnamefont {W.}~\bibnamefont {Yang}}, \bibinfo
  {author} {\bibfnamefont {C.}~\bibnamefont {Niu}}, \bibinfo {author}
  {\bibfnamefont {B.}~\bibnamefont {Chen}}, \bibinfo {author} {\bibfnamefont
  {C.}~\bibnamefont {Cao}},  \emph {et~al.},\ }\href@noop {} {\bibfield
  {journal} {\bibinfo  {journal} {Phys. Rev. B}\ }\textbf {\bibinfo {volume}
  {90}},\ \bibinfo {pages} {094520} (\bibinfo {year} {2014})}\BibitemShut
  {NoStop}%
\bibitem [{\citenamefont {Fulde}\ and\ \citenamefont
  {Ferrell}(1964)}]{fulde1964superconductivity}%
  \BibitemOpen
  \bibfield  {author} {\bibinfo {author} {\bibfnamefont {P.}~\bibnamefont
  {Fulde}}\ and\ \bibinfo {author} {\bibfnamefont {R.~A.}\ \bibnamefont
  {Ferrell}},\ }\href@noop {} {\bibfield  {journal} {\bibinfo  {journal} {Phys.
  Rev.}\ }\textbf {\bibinfo {volume} {135}},\ \bibinfo {pages} {A550} (\bibinfo
  {year} {1964})}\BibitemShut {NoStop}%
\bibitem [{\citenamefont {Larkin}\ and\ \citenamefont
  {Ovchinnikov}(1965)}]{larkin1965nonuniform}%
  \BibitemOpen
  \bibfield  {author} {\bibinfo {author} {\bibfnamefont {A.}~\bibnamefont
  {Larkin}}\ and\ \bibinfo {author} {\bibfnamefont {Y.~N.}\ \bibnamefont
  {Ovchinnikov}},\ }\href@noop {} {\bibfield  {journal} {\bibinfo  {journal}
  {Sov. Phys. JETP}\ }\textbf {\bibinfo {volume} {20}},\ \bibinfo {pages} {762}
  (\bibinfo {year} {1965})}\BibitemShut {NoStop}%
\bibitem [{\citenamefont {Kitagawa}\ \emph {et~al.}(2018)\citenamefont
  {Kitagawa}, \citenamefont {Nakamine}, \citenamefont {Ishida}, \citenamefont
  {Jeevan}, \citenamefont {Geibel},\ and\ \citenamefont
  {Steglich}}]{kitagawa2018evidence}%
  \BibitemOpen
  \bibfield  {author} {\bibinfo {author} {\bibfnamefont {S.}~\bibnamefont
  {Kitagawa}}, \bibinfo {author} {\bibfnamefont {G.}~\bibnamefont {Nakamine}},
  \bibinfo {author} {\bibfnamefont {K.}~\bibnamefont {Ishida}}, \bibinfo
  {author} {\bibfnamefont {H.}~\bibnamefont {Jeevan}}, \bibinfo {author}
  {\bibfnamefont {C.}~\bibnamefont {Geibel}}, \ and\ \bibinfo {author}
  {\bibfnamefont {F.}~\bibnamefont {Steglich}},\ }\href@noop {} {\bibfield
  {journal} {\bibinfo  {journal} {Phys. Rev. Lett.}\ }\textbf {\bibinfo
  {volume} {121}},\ \bibinfo {pages} {157004} (\bibinfo {year}
  {2018})}\BibitemShut {NoStop}%
\bibitem [{\citenamefont {Matsuda}\ and\ \citenamefont
  {Shimahara}(2007)}]{matsuda2007fulde}%
  \BibitemOpen
  \bibfield  {author} {\bibinfo {author} {\bibfnamefont {Y.}~\bibnamefont
  {Matsuda}}\ and\ \bibinfo {author} {\bibfnamefont {H.}~\bibnamefont
  {Shimahara}},\ }\href@noop {} {\bibfield  {journal} {\bibinfo  {journal} {J.
  Phys. Soc. Jpn.}\ }\textbf {\bibinfo {volume} {76}},\ \bibinfo {pages}
  {051005} (\bibinfo {year} {2007})}\BibitemShut {NoStop}%
\bibitem [{\citenamefont {Bianchi}\ \emph {et~al.}(2003)\citenamefont
  {Bianchi}, \citenamefont {Movshovich}, \citenamefont {Capan}, \citenamefont
  {Pagliuso},\ and\ \citenamefont {Sarrao}}]{bianchi2003possible}%
  \BibitemOpen
  \bibfield  {author} {\bibinfo {author} {\bibfnamefont {A.}~\bibnamefont
  {Bianchi}}, \bibinfo {author} {\bibfnamefont {R.}~\bibnamefont {Movshovich}},
  \bibinfo {author} {\bibfnamefont {C.}~\bibnamefont {Capan}}, \bibinfo
  {author} {\bibfnamefont {P.}~\bibnamefont {Pagliuso}}, \ and\ \bibinfo
  {author} {\bibfnamefont {J.}~\bibnamefont {Sarrao}},\ }\href@noop {}
  {\bibfield  {journal} {\bibinfo  {journal} {Phys. Rev. Lett.}\ }\textbf
  {\bibinfo {volume} {91}},\ \bibinfo {pages} {187004} (\bibinfo {year}
  {2003})}\BibitemShut {NoStop}%
\bibitem [{\citenamefont {Singleton}\ \emph {et~al.}(2000)\citenamefont
  {Singleton}, \citenamefont {Symington}, \citenamefont {Nam}, \citenamefont
  {Ardavan}, \citenamefont {Kurmoo},\ and\ \citenamefont
  {Day}}]{singleton2000observation}%
  \BibitemOpen
  \bibfield  {author} {\bibinfo {author} {\bibfnamefont {J.}~\bibnamefont
  {Singleton}}, \bibinfo {author} {\bibfnamefont {J.}~\bibnamefont
  {Symington}}, \bibinfo {author} {\bibfnamefont {M.}~\bibnamefont {Nam}},
  \bibinfo {author} {\bibfnamefont {A.}~\bibnamefont {Ardavan}}, \bibinfo
  {author} {\bibfnamefont {M.}~\bibnamefont {Kurmoo}}, \ and\ \bibinfo {author}
  {\bibfnamefont {P.}~\bibnamefont {Day}},\ }\href@noop {} {\bibfield
  {journal} {\bibinfo  {journal} {J. Phys.: Condens. Matter}\ }\textbf
  {\bibinfo {volume} {12}},\ \bibinfo {pages} {L641} (\bibinfo {year}
  {2000})}\BibitemShut {NoStop}%
\bibitem [{\citenamefont {Lortz}\ \emph {et~al.}(2007)\citenamefont {Lortz},
  \citenamefont {Wang}, \citenamefont {Demuer}, \citenamefont {B{\"o}ttger},
  \citenamefont {Bergk}, \citenamefont {Zwicknagl}, \citenamefont {Nakazawa},\
  and\ \citenamefont {Wosnitza}}]{lortz2007calorimetric}%
  \BibitemOpen
  \bibfield  {author} {\bibinfo {author} {\bibfnamefont {R.}~\bibnamefont
  {Lortz}}, \bibinfo {author} {\bibfnamefont {Y.}~\bibnamefont {Wang}},
  \bibinfo {author} {\bibfnamefont {A.}~\bibnamefont {Demuer}}, \bibinfo
  {author} {\bibfnamefont {P.}~\bibnamefont {B{\"o}ttger}}, \bibinfo {author}
  {\bibfnamefont {B.}~\bibnamefont {Bergk}}, \bibinfo {author} {\bibfnamefont
  {G.}~\bibnamefont {Zwicknagl}}, \bibinfo {author} {\bibfnamefont
  {Y.}~\bibnamefont {Nakazawa}}, \ and\ \bibinfo {author} {\bibfnamefont
  {J.}~\bibnamefont {Wosnitza}},\ }\href@noop {} {\bibfield  {journal}
  {\bibinfo  {journal} {Phys. Rev. Lett.}\ }\textbf {\bibinfo {volume} {99}},\
  \bibinfo {pages} {187002} (\bibinfo {year} {2007})}\BibitemShut {NoStop}%
\bibitem [{\citenamefont {Bergk}\ \emph {et~al.}(2011)\citenamefont {Bergk},
  \citenamefont {Demuer}, \citenamefont {Sheikin}, \citenamefont {Wang},
  \citenamefont {Wosnitza}, \citenamefont {Nakazawa},\ and\ \citenamefont
  {Lortz}}]{bergk2011magnetic}%
  \BibitemOpen
  \bibfield  {author} {\bibinfo {author} {\bibfnamefont {B.}~\bibnamefont
  {Bergk}}, \bibinfo {author} {\bibfnamefont {A.}~\bibnamefont {Demuer}},
  \bibinfo {author} {\bibfnamefont {I.}~\bibnamefont {Sheikin}}, \bibinfo
  {author} {\bibfnamefont {Y.}~\bibnamefont {Wang}}, \bibinfo {author}
  {\bibfnamefont {J.}~\bibnamefont {Wosnitza}}, \bibinfo {author}
  {\bibfnamefont {Y.}~\bibnamefont {Nakazawa}}, \ and\ \bibinfo {author}
  {\bibfnamefont {R.}~\bibnamefont {Lortz}},\ }\href@noop {} {\bibfield
  {journal} {\bibinfo  {journal} {Phys. Rev. B}\ }\textbf {\bibinfo {volume}
  {83}},\ \bibinfo {pages} {064506} (\bibinfo {year} {2011})}\BibitemShut
  {NoStop}%
\bibitem [{\citenamefont {Beyer}\ \emph {et~al.}(2012)\citenamefont {Beyer},
  \citenamefont {Bergk}, \citenamefont {Yasin}, \citenamefont {Schlueter},\
  and\ \citenamefont {Wosnitza}}]{beyer2012angle}%
  \BibitemOpen
  \bibfield  {author} {\bibinfo {author} {\bibfnamefont {R.}~\bibnamefont
  {Beyer}}, \bibinfo {author} {\bibfnamefont {B.}~\bibnamefont {Bergk}},
  \bibinfo {author} {\bibfnamefont {S.}~\bibnamefont {Yasin}}, \bibinfo
  {author} {\bibfnamefont {J.}~\bibnamefont {Schlueter}}, \ and\ \bibinfo
  {author} {\bibfnamefont {J.}~\bibnamefont {Wosnitza}},\ }\href@noop {}
  {\bibfield  {journal} {\bibinfo  {journal} {Phys. Rev. Lett.}\ }\textbf
  {\bibinfo {volume} {109}},\ \bibinfo {pages} {027003} (\bibinfo {year}
  {2012})}\BibitemShut {NoStop}%
\bibitem [{\citenamefont {Koutroulakis}\ \emph {et~al.}(2016)\citenamefont
  {Koutroulakis}, \citenamefont {K{\"u}hne}, \citenamefont {Schlueter},
  \citenamefont {Wosnitza},\ and\ \citenamefont
  {Brown}}]{koutroulakis2016microscopic}%
  \BibitemOpen
  \bibfield  {author} {\bibinfo {author} {\bibfnamefont {G.}~\bibnamefont
  {Koutroulakis}}, \bibinfo {author} {\bibfnamefont {H.}~\bibnamefont
  {K{\"u}hne}}, \bibinfo {author} {\bibfnamefont {J.}~\bibnamefont
  {Schlueter}}, \bibinfo {author} {\bibfnamefont {J.}~\bibnamefont {Wosnitza}},
  \ and\ \bibinfo {author} {\bibfnamefont {S.}~\bibnamefont {Brown}},\
  }\href@noop {} {\bibfield  {journal} {\bibinfo  {journal} {Phys. Rev. Lett.}\
  }\textbf {\bibinfo {volume} {116}},\ \bibinfo {pages} {067003} (\bibinfo
  {year} {2016})}\BibitemShut {NoStop}%
\bibitem [{\citenamefont {Tsuchiya}\ \emph {et~al.}(2015)\citenamefont
  {Tsuchiya}, \citenamefont {Yamada}, \citenamefont {Sugii}, \citenamefont
  {Graf}, \citenamefont {Brooks}, \citenamefont {Terashima},\ and\
  \citenamefont {Uji}}]{tsuchiya2015phase}%
  \BibitemOpen
  \bibfield  {author} {\bibinfo {author} {\bibfnamefont {S.}~\bibnamefont
  {Tsuchiya}}, \bibinfo {author} {\bibfnamefont {J.-i.}\ \bibnamefont
  {Yamada}}, \bibinfo {author} {\bibfnamefont {K.}~\bibnamefont {Sugii}},
  \bibinfo {author} {\bibfnamefont {D.}~\bibnamefont {Graf}}, \bibinfo {author}
  {\bibfnamefont {J.~S.}\ \bibnamefont {Brooks}}, \bibinfo {author}
  {\bibfnamefont {T.}~\bibnamefont {Terashima}}, \ and\ \bibinfo {author}
  {\bibfnamefont {S.}~\bibnamefont {Uji}},\ }\href@noop {} {\bibfield
  {journal} {\bibinfo  {journal} {J. Phys. Soc. Jpn.}\ }\textbf {\bibinfo
  {volume} {84}},\ \bibinfo {pages} {034703} (\bibinfo {year}
  {2015})}\BibitemShut {NoStop}%
\bibitem [{\citenamefont {Agosta}\ \emph {et~al.}(2012)\citenamefont {Agosta},
  \citenamefont {Jin}, \citenamefont {Coniglio}, \citenamefont {Smith},
  \citenamefont {Cho}, \citenamefont {Stroe}, \citenamefont {Martin},
  \citenamefont {Tozer}, \citenamefont {Murphy}, \citenamefont {Palm} \emph
  {et~al.}}]{agosta2012experimental}%
  \BibitemOpen
  \bibfield  {author} {\bibinfo {author} {\bibfnamefont {C.}~\bibnamefont
  {Agosta}}, \bibinfo {author} {\bibfnamefont {J.}~\bibnamefont {Jin}},
  \bibinfo {author} {\bibfnamefont {W.}~\bibnamefont {Coniglio}}, \bibinfo
  {author} {\bibfnamefont {B.}~\bibnamefont {Smith}}, \bibinfo {author}
  {\bibfnamefont {K.}~\bibnamefont {Cho}}, \bibinfo {author} {\bibfnamefont
  {I.}~\bibnamefont {Stroe}}, \bibinfo {author} {\bibfnamefont
  {C.}~\bibnamefont {Martin}}, \bibinfo {author} {\bibfnamefont
  {S.}~\bibnamefont {Tozer}}, \bibinfo {author} {\bibfnamefont
  {T.}~\bibnamefont {Murphy}}, \bibinfo {author} {\bibfnamefont
  {E.}~\bibnamefont {Palm}},  \emph {et~al.},\ }\href@noop {} {\bibfield
  {journal} {\bibinfo  {journal} {Phys. Rev. B}\ }\textbf {\bibinfo {volume}
  {85}},\ \bibinfo {pages} {214514} (\bibinfo {year} {2012})}\BibitemShut
  {NoStop}%
\bibitem [{\citenamefont {Wright}\ \emph {et~al.}(2011)\citenamefont {Wright},
  \citenamefont {Green}, \citenamefont {Kuhns}, \citenamefont {Reyes},
  \citenamefont {Brooks}, \citenamefont {Schlueter}, \citenamefont {Kato},
  \citenamefont {Yamamoto}, \citenamefont {Kobayashi},\ and\ \citenamefont
  {Brown}}]{wright2011zeeman}%
  \BibitemOpen
  \bibfield  {author} {\bibinfo {author} {\bibfnamefont {J.}~\bibnamefont
  {Wright}}, \bibinfo {author} {\bibfnamefont {E.}~\bibnamefont {Green}},
  \bibinfo {author} {\bibfnamefont {P.}~\bibnamefont {Kuhns}}, \bibinfo
  {author} {\bibfnamefont {A.}~\bibnamefont {Reyes}}, \bibinfo {author}
  {\bibfnamefont {J.}~\bibnamefont {Brooks}}, \bibinfo {author} {\bibfnamefont
  {J.}~\bibnamefont {Schlueter}}, \bibinfo {author} {\bibfnamefont
  {R.}~\bibnamefont {Kato}}, \bibinfo {author} {\bibfnamefont {H.}~\bibnamefont
  {Yamamoto}}, \bibinfo {author} {\bibfnamefont {M.}~\bibnamefont {Kobayashi}},
  \ and\ \bibinfo {author} {\bibfnamefont {S.}~\bibnamefont {Brown}},\
  }\href@noop {} {\bibfield  {journal} {\bibinfo  {journal} {Phys. Rev. Lett.}\
  }\textbf {\bibinfo {volume} {107}},\ \bibinfo {pages} {087002} (\bibinfo
  {year} {2011})}\BibitemShut {NoStop}%
\bibitem [{\citenamefont {Mayaffre}\ \emph {et~al.}(2014)\citenamefont
  {Mayaffre}, \citenamefont {Kr{\"a}mer}, \citenamefont {Horvati{\'c}},
  \citenamefont {Berthier}, \citenamefont {Miyagawa}, \citenamefont {Kanoda},\
  and\ \citenamefont {Mitrovi{\'c}}}]{mayaffre2014evidence}%
  \BibitemOpen
  \bibfield  {author} {\bibinfo {author} {\bibfnamefont {H.}~\bibnamefont
  {Mayaffre}}, \bibinfo {author} {\bibfnamefont {S.}~\bibnamefont
  {Kr{\"a}mer}}, \bibinfo {author} {\bibfnamefont {M.}~\bibnamefont
  {Horvati{\'c}}}, \bibinfo {author} {\bibfnamefont {C.}~\bibnamefont
  {Berthier}}, \bibinfo {author} {\bibfnamefont {K.}~\bibnamefont {Miyagawa}},
  \bibinfo {author} {\bibfnamefont {K.}~\bibnamefont {Kanoda}}, \ and\ \bibinfo
  {author} {\bibfnamefont {V.}~\bibnamefont {Mitrovi{\'c}}},\ }\href@noop {}
  {\bibfield  {journal} {\bibinfo  {journal} {Nat. Phys.}\ }\textbf {\bibinfo
  {volume} {10}},\ \bibinfo {pages} {928} (\bibinfo {year} {2014})}\BibitemShut
  {NoStop}%
\bibitem [{\citenamefont {Cho}\ \emph {et~al.}(2009)\citenamefont {Cho},
  \citenamefont {Smith}, \citenamefont {Coniglio}, \citenamefont {Winter},
  \citenamefont {Agosta},\ and\ \citenamefont {Schlueter}}]{cho2009upper}%
  \BibitemOpen
  \bibfield  {author} {\bibinfo {author} {\bibfnamefont {K.}~\bibnamefont
  {Cho}}, \bibinfo {author} {\bibfnamefont {B.}~\bibnamefont {Smith}}, \bibinfo
  {author} {\bibfnamefont {W.}~\bibnamefont {Coniglio}}, \bibinfo {author}
  {\bibfnamefont {L.}~\bibnamefont {Winter}}, \bibinfo {author} {\bibfnamefont
  {C.}~\bibnamefont {Agosta}}, \ and\ \bibinfo {author} {\bibfnamefont
  {J.}~\bibnamefont {Schlueter}},\ }\href@noop {} {\bibfield  {journal}
  {\bibinfo  {journal} {Phys. Rev. B}\ }\textbf {\bibinfo {volume} {79}},\
  \bibinfo {pages} {220507} (\bibinfo {year} {2009})}\BibitemShut {NoStop}%
\bibitem [{\citenamefont {Coniglio}\ \emph {et~al.}(2011)\citenamefont
  {Coniglio}, \citenamefont {Winter}, \citenamefont {Cho}, \citenamefont
  {Agosta}, \citenamefont {Fravel},\ and\ \citenamefont
  {Montgomery}}]{coniglio2011superconducting}%
  \BibitemOpen
  \bibfield  {author} {\bibinfo {author} {\bibfnamefont {W.~A.}\ \bibnamefont
  {Coniglio}}, \bibinfo {author} {\bibfnamefont {L.~E.}\ \bibnamefont
  {Winter}}, \bibinfo {author} {\bibfnamefont {K.}~\bibnamefont {Cho}},
  \bibinfo {author} {\bibfnamefont {C.}~\bibnamefont {Agosta}}, \bibinfo
  {author} {\bibfnamefont {B.}~\bibnamefont {Fravel}}, \ and\ \bibinfo {author}
  {\bibfnamefont {L.}~\bibnamefont {Montgomery}},\ }\href@noop {} {\bibfield
  {journal} {\bibinfo  {journal} {Phys. Rev. B}\ }\textbf {\bibinfo {volume}
  {83}},\ \bibinfo {pages} {224507} (\bibinfo {year} {2011})}\BibitemShut
  {NoStop}%
\bibitem [{\citenamefont {Ok}\ \emph {et~al.}(2020)\citenamefont {Ok},
  \citenamefont {Kwon}, \citenamefont {Kohama}, \citenamefont {You},
  \citenamefont {Park}, \citenamefont {Kim}, \citenamefont {Jo}, \citenamefont
  {Choi}, \citenamefont {Kindo}, \citenamefont {Kang} \emph
  {et~al.}}]{ok2020observation}%
  \BibitemOpen
  \bibfield  {author} {\bibinfo {author} {\bibfnamefont {J.~M.}\ \bibnamefont
  {Ok}}, \bibinfo {author} {\bibfnamefont {C.~I.}\ \bibnamefont {Kwon}},
  \bibinfo {author} {\bibfnamefont {Y.}~\bibnamefont {Kohama}}, \bibinfo
  {author} {\bibfnamefont {J.~S.}\ \bibnamefont {You}}, \bibinfo {author}
  {\bibfnamefont {S.~K.}\ \bibnamefont {Park}}, \bibinfo {author}
  {\bibfnamefont {J.-h.}\ \bibnamefont {Kim}}, \bibinfo {author} {\bibfnamefont
  {Y.}~\bibnamefont {Jo}}, \bibinfo {author} {\bibfnamefont {E.}~\bibnamefont
  {Choi}}, \bibinfo {author} {\bibfnamefont {K.}~\bibnamefont {Kindo}},
  \bibinfo {author} {\bibfnamefont {W.}~\bibnamefont {Kang}},  \emph {et~al.},\
  }\href@noop {} {\bibfield  {journal} {\bibinfo  {journal} {Phys. Rev. B}\
  }\textbf {\bibinfo {volume} {101}},\ \bibinfo {pages} {224509} (\bibinfo
  {year} {2020})}\BibitemShut {NoStop}%
\bibitem [{\citenamefont {Kasahara}\ \emph {et~al.}(2020)\citenamefont
  {Kasahara}, \citenamefont {Sato}, \citenamefont {Licciardello}, \citenamefont
  {{\v{C}}ulo}, \citenamefont {Arsenijevi{\'c}}, \citenamefont {Ottenbros},
  \citenamefont {Tominaga}, \citenamefont {B{\"o}ker}, \citenamefont {Eremin},
  \citenamefont {Shibauchi} \emph {et~al.}}]{kasahara2020evidence}%
  \BibitemOpen
  \bibfield  {author} {\bibinfo {author} {\bibfnamefont {S.}~\bibnamefont
  {Kasahara}}, \bibinfo {author} {\bibfnamefont {Y.}~\bibnamefont {Sato}},
  \bibinfo {author} {\bibfnamefont {S.}~\bibnamefont {Licciardello}}, \bibinfo
  {author} {\bibfnamefont {M.}~\bibnamefont {{\v{C}}ulo}}, \bibinfo {author}
  {\bibfnamefont {S.}~\bibnamefont {Arsenijevi{\'c}}}, \bibinfo {author}
  {\bibfnamefont {T.}~\bibnamefont {Ottenbros}}, \bibinfo {author}
  {\bibfnamefont {T.}~\bibnamefont {Tominaga}}, \bibinfo {author}
  {\bibfnamefont {J.}~\bibnamefont {B{\"o}ker}}, \bibinfo {author}
  {\bibfnamefont {I.}~\bibnamefont {Eremin}}, \bibinfo {author} {\bibfnamefont
  {T.}~\bibnamefont {Shibauchi}},  \emph {et~al.},\ }\href@noop {} {\bibfield
  {journal} {\bibinfo  {journal} {Phys. Rev. Lett.}\ }\textbf {\bibinfo
  {volume} {124}},\ \bibinfo {pages} {107001} (\bibinfo {year}
  {2020})}\BibitemShut {NoStop}%
\bibitem [{\citenamefont {Kasahara}\ \emph {et~al.}(2014)\citenamefont
  {Kasahara}, \citenamefont {Watashige}, \citenamefont {Hanaguri},
  \citenamefont {Kohsaka}, \citenamefont {Yamashita}, \citenamefont
  {Shimoyama}, \citenamefont {Mizukami}, \citenamefont {Endo}, \citenamefont
  {Ikeda}, \citenamefont {Aoyama} \emph {et~al.}}]{kasahara2014field}%
  \BibitemOpen
  \bibfield  {author} {\bibinfo {author} {\bibfnamefont {S.}~\bibnamefont
  {Kasahara}}, \bibinfo {author} {\bibfnamefont {T.}~\bibnamefont {Watashige}},
  \bibinfo {author} {\bibfnamefont {T.}~\bibnamefont {Hanaguri}}, \bibinfo
  {author} {\bibfnamefont {Y.}~\bibnamefont {Kohsaka}}, \bibinfo {author}
  {\bibfnamefont {T.}~\bibnamefont {Yamashita}}, \bibinfo {author}
  {\bibfnamefont {Y.}~\bibnamefont {Shimoyama}}, \bibinfo {author}
  {\bibfnamefont {Y.}~\bibnamefont {Mizukami}}, \bibinfo {author}
  {\bibfnamefont {R.}~\bibnamefont {Endo}}, \bibinfo {author} {\bibfnamefont
  {H.}~\bibnamefont {Ikeda}}, \bibinfo {author} {\bibfnamefont
  {K.}~\bibnamefont {Aoyama}},  \emph {et~al.},\ }\href@noop {} {\bibfield
  {journal} {\bibinfo  {journal} {Proc. Natl. Acad. Sci. U. S. A.}\ }\textbf
  {\bibinfo {volume} {111}},\ \bibinfo {pages} {16309} (\bibinfo {year}
  {2014})}\BibitemShut {NoStop}%
\bibitem [{\citenamefont {Cho}\ \emph {et~al.}(2017)\citenamefont {Cho},
  \citenamefont {Yang}, \citenamefont {Yuan}, \citenamefont {Shen},
  \citenamefont {Wolf},\ and\ \citenamefont {Lortz}}]{cho2017thermodynamic}%
  \BibitemOpen
  \bibfield  {author} {\bibinfo {author} {\bibfnamefont {C.-w.}\ \bibnamefont
  {Cho}}, \bibinfo {author} {\bibfnamefont {J.~H.}\ \bibnamefont {Yang}},
  \bibinfo {author} {\bibfnamefont {N.~F.}\ \bibnamefont {Yuan}}, \bibinfo
  {author} {\bibfnamefont {J.}~\bibnamefont {Shen}}, \bibinfo {author}
  {\bibfnamefont {T.}~\bibnamefont {Wolf}}, \ and\ \bibinfo {author}
  {\bibfnamefont {R.}~\bibnamefont {Lortz}},\ }\href@noop {} {\bibfield
  {journal} {\bibinfo  {journal} {Phys. Rev. Lett.}\ }\textbf {\bibinfo
  {volume} {119}},\ \bibinfo {pages} {217002} (\bibinfo {year}
  {2017})}\BibitemShut {NoStop}%
\bibitem [{\citenamefont {Kenzelmann}\ \emph {et~al.}(2008)\citenamefont
  {Kenzelmann}, \citenamefont {Str{\"a}ssle}, \citenamefont {Niedermayer},
  \citenamefont {Sigrist}, \citenamefont {Padmanabhan}, \citenamefont
  {Zolliker}, \citenamefont {Bianchi}, \citenamefont {Movshovich},
  \citenamefont {Bauer}, \citenamefont {Sarrao} \emph
  {et~al.}}]{kenzelmann2008coupled}%
  \BibitemOpen
  \bibfield  {author} {\bibinfo {author} {\bibfnamefont {M.}~\bibnamefont
  {Kenzelmann}}, \bibinfo {author} {\bibfnamefont {T.}~\bibnamefont
  {Str{\"a}ssle}}, \bibinfo {author} {\bibfnamefont {C.}~\bibnamefont
  {Niedermayer}}, \bibinfo {author} {\bibfnamefont {M.}~\bibnamefont
  {Sigrist}}, \bibinfo {author} {\bibfnamefont {B.}~\bibnamefont
  {Padmanabhan}}, \bibinfo {author} {\bibfnamefont {M.}~\bibnamefont
  {Zolliker}}, \bibinfo {author} {\bibfnamefont {A.}~\bibnamefont {Bianchi}},
  \bibinfo {author} {\bibfnamefont {R.}~\bibnamefont {Movshovich}}, \bibinfo
  {author} {\bibfnamefont {E.~D.}\ \bibnamefont {Bauer}}, \bibinfo {author}
  {\bibfnamefont {J.~L.}\ \bibnamefont {Sarrao}},  \emph {et~al.},\ }\href@noop
  {} {\bibfield  {journal} {\bibinfo  {journal} {Science}\ }\textbf {\bibinfo
  {volume} {321}},\ \bibinfo {pages} {1652} (\bibinfo {year}
  {2008})}\BibitemShut {NoStop}%
\bibitem [{\citenamefont {Kenzelmann}\ \emph {et~al.}(2010)\citenamefont
  {Kenzelmann}, \citenamefont {Gerber}, \citenamefont {Egetenmeyer},
  \citenamefont {Gavilano}, \citenamefont {Str{\"a}ssle}, \citenamefont
  {Bianchi}, \citenamefont {Ressouche}, \citenamefont {Movshovich},
  \citenamefont {Bauer}, \citenamefont {Sarrao} \emph
  {et~al.}}]{kenzelmann2010evidence}%
  \BibitemOpen
  \bibfield  {author} {\bibinfo {author} {\bibfnamefont {M.}~\bibnamefont
  {Kenzelmann}}, \bibinfo {author} {\bibfnamefont {S.}~\bibnamefont {Gerber}},
  \bibinfo {author} {\bibfnamefont {N.}~\bibnamefont {Egetenmeyer}}, \bibinfo
  {author} {\bibfnamefont {J.}~\bibnamefont {Gavilano}}, \bibinfo {author}
  {\bibfnamefont {T.}~\bibnamefont {Str{\"a}ssle}}, \bibinfo {author}
  {\bibfnamefont {A.}~\bibnamefont {Bianchi}}, \bibinfo {author} {\bibfnamefont
  {E.}~\bibnamefont {Ressouche}}, \bibinfo {author} {\bibfnamefont
  {R.}~\bibnamefont {Movshovich}}, \bibinfo {author} {\bibfnamefont
  {E.}~\bibnamefont {Bauer}}, \bibinfo {author} {\bibfnamefont
  {J.}~\bibnamefont {Sarrao}},  \emph {et~al.},\ }\href@noop {} {\bibfield
  {journal} {\bibinfo  {journal} {Phys. Rev. Lett.}\ }\textbf {\bibinfo
  {volume} {104}},\ \bibinfo {pages} {127001} (\bibinfo {year}
  {2010})}\BibitemShut {NoStop}%
\bibitem [{\citenamefont {Hsu}\ \emph {et~al.}(2008)\citenamefont {Hsu},
  \citenamefont {Luo}, \citenamefont {Yeh}, \citenamefont {Chen}, \citenamefont
  {Huang}, \citenamefont {Wu}, \citenamefont {Lee}, \citenamefont {Huang},
  \citenamefont {Chu}, \citenamefont {Yan} \emph
  {et~al.}}]{hsu2008superconductivity}%
  \BibitemOpen
  \bibfield  {author} {\bibinfo {author} {\bibfnamefont {F.-C.}\ \bibnamefont
  {Hsu}}, \bibinfo {author} {\bibfnamefont {J.-Y.}\ \bibnamefont {Luo}},
  \bibinfo {author} {\bibfnamefont {K.-W.}\ \bibnamefont {Yeh}}, \bibinfo
  {author} {\bibfnamefont {T.-K.}\ \bibnamefont {Chen}}, \bibinfo {author}
  {\bibfnamefont {T.-W.}\ \bibnamefont {Huang}}, \bibinfo {author}
  {\bibfnamefont {P.~M.}\ \bibnamefont {Wu}}, \bibinfo {author} {\bibfnamefont
  {Y.-C.}\ \bibnamefont {Lee}}, \bibinfo {author} {\bibfnamefont {Y.-L.}\
  \bibnamefont {Huang}}, \bibinfo {author} {\bibfnamefont {Y.-Y.}\ \bibnamefont
  {Chu}}, \bibinfo {author} {\bibfnamefont {D.-C.}\ \bibnamefont {Yan}},  \emph
  {et~al.},\ }\href@noop {} {\bibfield  {journal} {\bibinfo  {journal} {Proc.
  Natl. Acad. Sci. U. S. A.}\ }\textbf {\bibinfo {volume} {105}},\ \bibinfo
  {pages} {14262} (\bibinfo {year} {2008})}\BibitemShut {NoStop}%
\bibitem [{\citenamefont {Watson}\ \emph {et~al.}(2015)\citenamefont {Watson},
  \citenamefont {Yamashita}, \citenamefont {Kasahara}, \citenamefont {Knafo},
  \citenamefont {Nardone}, \citenamefont {B{\'e}ard}, \citenamefont {Hardy},
  \citenamefont {McCollam}, \citenamefont {Narayanan}, \citenamefont {Blake}
  \emph {et~al.}}]{watson2015dichotomy}%
  \BibitemOpen
  \bibfield  {author} {\bibinfo {author} {\bibfnamefont {M.}~\bibnamefont
  {Watson}}, \bibinfo {author} {\bibfnamefont {T.}~\bibnamefont {Yamashita}},
  \bibinfo {author} {\bibfnamefont {S.}~\bibnamefont {Kasahara}}, \bibinfo
  {author} {\bibfnamefont {W.}~\bibnamefont {Knafo}}, \bibinfo {author}
  {\bibfnamefont {M.}~\bibnamefont {Nardone}}, \bibinfo {author} {\bibfnamefont
  {J.}~\bibnamefont {B{\'e}ard}}, \bibinfo {author} {\bibfnamefont
  {F.}~\bibnamefont {Hardy}}, \bibinfo {author} {\bibfnamefont
  {A.}~\bibnamefont {McCollam}}, \bibinfo {author} {\bibfnamefont
  {A.}~\bibnamefont {Narayanan}}, \bibinfo {author} {\bibfnamefont
  {S.}~\bibnamefont {Blake}},  \emph {et~al.},\ }\href@noop {} {\bibfield
  {journal} {\bibinfo  {journal} {Phys. Rev. Lett.}\ }\textbf {\bibinfo
  {volume} {115}},\ \bibinfo {pages} {027006} (\bibinfo {year}
  {2015})}\BibitemShut {NoStop}%
\bibitem [{\citenamefont {Hosoi}\ \emph {et~al.}(2016)\citenamefont {Hosoi},
  \citenamefont {Matsuura}, \citenamefont {Ishida}, \citenamefont {Wang},
  \citenamefont {Mizukami}, \citenamefont {Watashige}, \citenamefont
  {Kasahara}, \citenamefont {Matsuda},\ and\ \citenamefont
  {Shibauchi}}]{hosoi2016nematic}%
  \BibitemOpen
  \bibfield  {author} {\bibinfo {author} {\bibfnamefont {S.}~\bibnamefont
  {Hosoi}}, \bibinfo {author} {\bibfnamefont {K.}~\bibnamefont {Matsuura}},
  \bibinfo {author} {\bibfnamefont {K.}~\bibnamefont {Ishida}}, \bibinfo
  {author} {\bibfnamefont {H.}~\bibnamefont {Wang}}, \bibinfo {author}
  {\bibfnamefont {Y.}~\bibnamefont {Mizukami}}, \bibinfo {author}
  {\bibfnamefont {T.}~\bibnamefont {Watashige}}, \bibinfo {author}
  {\bibfnamefont {S.}~\bibnamefont {Kasahara}}, \bibinfo {author}
  {\bibfnamefont {Y.}~\bibnamefont {Matsuda}}, \ and\ \bibinfo {author}
  {\bibfnamefont {T.}~\bibnamefont {Shibauchi}},\ }\href@noop {} {\bibfield
  {journal} {\bibinfo  {journal} {Proc. Natl. Acad. Sci. U. S. A.}\ }\textbf
  {\bibinfo {volume} {113}},\ \bibinfo {pages} {8139} (\bibinfo {year}
  {2016})}\BibitemShut {NoStop}%
\bibitem [{\citenamefont {Sprau}\ \emph {et~al.}(2017)\citenamefont {Sprau},
  \citenamefont {Kostin}, \citenamefont {Kreisel}, \citenamefont {B{\"o}hmer},
  \citenamefont {Taufour}, \citenamefont {Canfield}, \citenamefont {Mukherjee},
  \citenamefont {Hirschfeld}, \citenamefont {Andersen},\ and\ \citenamefont
  {Davis}}]{sprau2017discovery}%
  \BibitemOpen
  \bibfield  {author} {\bibinfo {author} {\bibfnamefont {P.~O.}\ \bibnamefont
  {Sprau}}, \bibinfo {author} {\bibfnamefont {A.}~\bibnamefont {Kostin}},
  \bibinfo {author} {\bibfnamefont {A.}~\bibnamefont {Kreisel}}, \bibinfo
  {author} {\bibfnamefont {A.~E.}\ \bibnamefont {B{\"o}hmer}}, \bibinfo
  {author} {\bibfnamefont {V.}~\bibnamefont {Taufour}}, \bibinfo {author}
  {\bibfnamefont {P.~C.}\ \bibnamefont {Canfield}}, \bibinfo {author}
  {\bibfnamefont {S.}~\bibnamefont {Mukherjee}}, \bibinfo {author}
  {\bibfnamefont {P.~J.}\ \bibnamefont {Hirschfeld}}, \bibinfo {author}
  {\bibfnamefont {B.~M.}\ \bibnamefont {Andersen}}, \ and\ \bibinfo {author}
  {\bibfnamefont {J.~S.}\ \bibnamefont {Davis}},\ }\href@noop {} {\bibfield
  {journal} {\bibinfo  {journal} {Science}\ }\textbf {\bibinfo {volume}
  {357}},\ \bibinfo {pages} {75} (\bibinfo {year} {2017})}\BibitemShut
  {NoStop}%
\bibitem [{\citenamefont {Aslamazov}(1969)}]{aslamazov1969influence}%
  \BibitemOpen
  \bibfield  {author} {\bibinfo {author} {\bibfnamefont {L.}~\bibnamefont
  {Aslamazov}},\ }\href@noop {} {\bibfield  {journal} {\bibinfo  {journal}
  {Sov. Phys. JETP}\ }\textbf {\bibinfo {volume} {28}},\ \bibinfo {pages} {773}
  (\bibinfo {year} {1969})}\BibitemShut {NoStop}%
\bibitem [{\citenamefont {Takada}(1970)}]{takada1970superconductivity}%
  \BibitemOpen
  \bibfield  {author} {\bibinfo {author} {\bibfnamefont {S.}~\bibnamefont
  {Takada}},\ }\href@noop {} {\bibfield  {journal} {\bibinfo  {journal} {Prog.
  Theor. Phys.}\ }\textbf {\bibinfo {volume} {43}},\ \bibinfo {pages} {27}
  (\bibinfo {year} {1970})}\BibitemShut {NoStop}%
\bibitem [{\citenamefont {Sun}\ \emph {et~al.}(2016)\citenamefont {Sun},
  \citenamefont {Pyon},\ and\ \citenamefont {Tamegai}}]{sun2016electron}%
  \BibitemOpen
  \bibfield  {author} {\bibinfo {author} {\bibfnamefont {Y.}~\bibnamefont
  {Sun}}, \bibinfo {author} {\bibfnamefont {S.}~\bibnamefont {Pyon}}, \ and\
  \bibinfo {author} {\bibfnamefont {T.}~\bibnamefont {Tamegai}},\ }\href@noop
  {} {\bibfield  {journal} {\bibinfo  {journal} {Phys. Rev. B}\ }\textbf
  {\bibinfo {volume} {93}},\ \bibinfo {pages} {104502} (\bibinfo {year}
  {2016})}\BibitemShut {NoStop}%
\bibitem [{\citenamefont {B{\"o}hmer}\ \emph {et~al.}(2016)\citenamefont
  {B{\"o}hmer}, \citenamefont {Taufour}, \citenamefont {Straszheim},
  \citenamefont {Wolf},\ and\ \citenamefont {Canfield}}]{bohmer2016variation}%
  \BibitemOpen
  \bibfield  {author} {\bibinfo {author} {\bibfnamefont {A.}~\bibnamefont
  {B{\"o}hmer}}, \bibinfo {author} {\bibfnamefont {V.}~\bibnamefont {Taufour}},
  \bibinfo {author} {\bibfnamefont {W.}~\bibnamefont {Straszheim}}, \bibinfo
  {author} {\bibfnamefont {T.}~\bibnamefont {Wolf}}, \ and\ \bibinfo {author}
  {\bibfnamefont {P.}~\bibnamefont {Canfield}},\ }\href@noop {} {\bibfield
  {journal} {\bibinfo  {journal} {Phys. Rev. B}\ }\textbf {\bibinfo {volume}
  {94}},\ \bibinfo {pages} {024526} (\bibinfo {year} {2016})}\BibitemShut
  {NoStop}%
\bibitem [{\citenamefont {McQueen}\ \emph {et~al.}(2009)\citenamefont
  {McQueen}, \citenamefont {Williams}, \citenamefont {Stephens}, \citenamefont
  {Tao}, \citenamefont {Zhu}, \citenamefont {Ksenofontov}, \citenamefont
  {Casper}, \citenamefont {Felser},\ and\ \citenamefont
  {Cava}}]{mcqueen2009tetragonal}%
  \BibitemOpen
  \bibfield  {author} {\bibinfo {author} {\bibfnamefont {T.}~\bibnamefont
  {McQueen}}, \bibinfo {author} {\bibfnamefont {A.}~\bibnamefont {Williams}},
  \bibinfo {author} {\bibfnamefont {P.}~\bibnamefont {Stephens}}, \bibinfo
  {author} {\bibfnamefont {J.}~\bibnamefont {Tao}}, \bibinfo {author}
  {\bibfnamefont {Y.}~\bibnamefont {Zhu}}, \bibinfo {author} {\bibfnamefont
  {V.}~\bibnamefont {Ksenofontov}}, \bibinfo {author} {\bibfnamefont
  {F.}~\bibnamefont {Casper}}, \bibinfo {author} {\bibfnamefont
  {C.}~\bibnamefont {Felser}}, \ and\ \bibinfo {author} {\bibfnamefont {R.~J.}\
  \bibnamefont {Cava}},\ }\href@noop {} {\bibfield  {journal} {\bibinfo
  {journal} {Phys. Rev. Lett.}\ }\textbf {\bibinfo {volume} {103}},\ \bibinfo
  {pages} {057002} (\bibinfo {year} {2009})}\BibitemShut {NoStop}%
\bibitem [{sup()}]{supplement}%
  \BibitemOpen
  \href@noop {} {\ }\bibinfo {note} {See Supplemental Material at [] for the
  first derivative of \textit{$\rho$}--\textit{T}, temperature dependent
  resistivity at various fixed fields, the \textit{$B$}$_{\rm{c}2}$ determined
  by the criterion of the onset of the SC transition, the estimation of the
  \textit{$B$}$_{\rm{c}2}^{\parallel \rm{ab}}$(0 K), the coherence length,
  anisotropy, and Maki parameter, field dependence of the magnetic torque
  \textit{$\tau$}(\textit{$B$}), the determinations of the irreversibility
  field \textit{$B$}$_{\rm{irr}}$, the temperature dependent
  \textit{$B$}$_{\rm{c}2}$ determined by the resistivity (end point) and torque
  magnetometry}\BibitemShut {NoStop}%
\bibitem [{\citenamefont {Jiao}\ \emph {et~al.}(2017)\citenamefont {Jiao},
  \citenamefont {R{\"o}{\ss}ler}, \citenamefont {Koz}, \citenamefont {Schwarz},
  \citenamefont {Kasinathan}, \citenamefont {R{\"o}{\ss}ler},\ and\
  \citenamefont {Wirth}}]{jiao2017impurity}%
  \BibitemOpen
  \bibfield  {author} {\bibinfo {author} {\bibfnamefont {L.}~\bibnamefont
  {Jiao}}, \bibinfo {author} {\bibfnamefont {S.}~\bibnamefont
  {R{\"o}{\ss}ler}}, \bibinfo {author} {\bibfnamefont {C.}~\bibnamefont {Koz}},
  \bibinfo {author} {\bibfnamefont {U.}~\bibnamefont {Schwarz}}, \bibinfo
  {author} {\bibfnamefont {D.}~\bibnamefont {Kasinathan}}, \bibinfo {author}
  {\bibfnamefont {U.~K.}\ \bibnamefont {R{\"o}{\ss}ler}}, \ and\ \bibinfo
  {author} {\bibfnamefont {S.}~\bibnamefont {Wirth}},\ }\href@noop {}
  {\bibfield  {journal} {\bibinfo  {journal} {Phys. Rev. B}\ }\textbf {\bibinfo
  {volume} {96}},\ \bibinfo {pages} {094504} (\bibinfo {year}
  {2017})}\BibitemShut {NoStop}%
\bibitem [{\citenamefont {Terashima}\ \emph {et~al.}(2014)\citenamefont
  {Terashima}, \citenamefont {Kikugawa}, \citenamefont {Kiswandhi},
  \citenamefont {Choi}, \citenamefont {Brooks}, \citenamefont {Kasahara},
  \citenamefont {Watashige}, \citenamefont {Ikeda}, \citenamefont {Shibauchi},
  \citenamefont {Matsuda} \emph {et~al.}}]{terashima2014anomalous}%
  \BibitemOpen
  \bibfield  {author} {\bibinfo {author} {\bibfnamefont {T.}~\bibnamefont
  {Terashima}}, \bibinfo {author} {\bibfnamefont {N.}~\bibnamefont {Kikugawa}},
  \bibinfo {author} {\bibfnamefont {A.}~\bibnamefont {Kiswandhi}}, \bibinfo
  {author} {\bibfnamefont {E.-S.}\ \bibnamefont {Choi}}, \bibinfo {author}
  {\bibfnamefont {J.~S.}\ \bibnamefont {Brooks}}, \bibinfo {author}
  {\bibfnamefont {S.}~\bibnamefont {Kasahara}}, \bibinfo {author}
  {\bibfnamefont {T.}~\bibnamefont {Watashige}}, \bibinfo {author}
  {\bibfnamefont {H.}~\bibnamefont {Ikeda}}, \bibinfo {author} {\bibfnamefont
  {T.}~\bibnamefont {Shibauchi}}, \bibinfo {author} {\bibfnamefont
  {Y.}~\bibnamefont {Matsuda}},  \emph {et~al.},\ }\href@noop {} {\bibfield
  {journal} {\bibinfo  {journal} {Phys. Rev. B}\ }\textbf {\bibinfo {volume}
  {90}},\ \bibinfo {pages} {144517} (\bibinfo {year} {2014})}\BibitemShut
  {NoStop}%
\bibitem [{\citenamefont {Li}\ \emph {et~al.}(2007)\citenamefont {Li},
  \citenamefont {Checkelsky}, \citenamefont {Komiya}, \citenamefont {Ando},\
  and\ \citenamefont {Ong}}]{li2007low}%
  \BibitemOpen
  \bibfield  {author} {\bibinfo {author} {\bibfnamefont {L.}~\bibnamefont
  {Li}}, \bibinfo {author} {\bibfnamefont {J.}~\bibnamefont {Checkelsky}},
  \bibinfo {author} {\bibfnamefont {S.}~\bibnamefont {Komiya}}, \bibinfo
  {author} {\bibfnamefont {Y.}~\bibnamefont {Ando}}, \ and\ \bibinfo {author}
  {\bibfnamefont {N.}~\bibnamefont {Ong}},\ }\href@noop {} {\bibfield
  {journal} {\bibinfo  {journal} {Nat. Phys.}\ }\textbf {\bibinfo {volume}
  {3}},\ \bibinfo {pages} {311} (\bibinfo {year} {2007})}\BibitemShut {NoStop}%
\bibitem [{\citenamefont {Kasahara}\ \emph {et~al.}(2012)\citenamefont
  {Kasahara}, \citenamefont {Shi}, \citenamefont {Hashimoto}, \citenamefont
  {Tonegawa}, \citenamefont {Mizukami}, \citenamefont {Shibauchi},
  \citenamefont {Sugimoto}, \citenamefont {Fukuda}, \citenamefont {Terashima},
  \citenamefont {Nevidomskyy} \emph {et~al.}}]{kasahara2012electronic}%
  \BibitemOpen
  \bibfield  {author} {\bibinfo {author} {\bibfnamefont {S.}~\bibnamefont
  {Kasahara}}, \bibinfo {author} {\bibfnamefont {H.}~\bibnamefont {Shi}},
  \bibinfo {author} {\bibfnamefont {K.}~\bibnamefont {Hashimoto}}, \bibinfo
  {author} {\bibfnamefont {S.}~\bibnamefont {Tonegawa}}, \bibinfo {author}
  {\bibfnamefont {Y.}~\bibnamefont {Mizukami}}, \bibinfo {author}
  {\bibfnamefont {T.}~\bibnamefont {Shibauchi}}, \bibinfo {author}
  {\bibfnamefont {K.}~\bibnamefont {Sugimoto}}, \bibinfo {author}
  {\bibfnamefont {T.}~\bibnamefont {Fukuda}}, \bibinfo {author} {\bibfnamefont
  {T.}~\bibnamefont {Terashima}}, \bibinfo {author} {\bibfnamefont {A.~H.}\
  \bibnamefont {Nevidomskyy}},  \emph {et~al.},\ }\href@noop {} {\bibfield
  {journal} {\bibinfo  {journal} {Nature}\ }\textbf {\bibinfo {volume} {486}},\
  \bibinfo {pages} {382} (\bibinfo {year} {2012})}\BibitemShut {NoStop}%
\bibitem [{\citenamefont {Wang}\ \emph {et~al.}(2020)\citenamefont {Wang},
  \citenamefont {Zhang}, \citenamefont {Xu}, \citenamefont {Wang},
  \citenamefont {Jiang}, \citenamefont {Zhu}, \citenamefont {Wang},
  \citenamefont {Chu}, \citenamefont {Feng}, \citenamefont {Wang} \emph
  {et~al.}}]{wang2020strong}%
  \BibitemOpen
  \bibfield  {author} {\bibinfo {author} {\bibfnamefont {T.}~\bibnamefont
  {Wang}}, \bibinfo {author} {\bibfnamefont {C.}~\bibnamefont {Zhang}},
  \bibinfo {author} {\bibfnamefont {L.}~\bibnamefont {Xu}}, \bibinfo {author}
  {\bibfnamefont {J.}~\bibnamefont {Wang}}, \bibinfo {author} {\bibfnamefont
  {S.}~\bibnamefont {Jiang}}, \bibinfo {author} {\bibfnamefont
  {Z.}~\bibnamefont {Zhu}}, \bibinfo {author} {\bibfnamefont {Z.}~\bibnamefont
  {Wang}}, \bibinfo {author} {\bibfnamefont {J.}~\bibnamefont {Chu}}, \bibinfo
  {author} {\bibfnamefont {J.}~\bibnamefont {Feng}}, \bibinfo {author}
  {\bibfnamefont {L.}~\bibnamefont {Wang}},  \emph {et~al.},\ }\href@noop {}
  {\bibfield  {journal} {\bibinfo  {journal} {Sci. China-Phys. Mech. Astron.}\
  }\textbf {\bibinfo {volume} {63}},\ \bibinfo {pages} {227412} (\bibinfo
  {year} {2020})}\BibitemShut {NoStop}%
\bibitem [{\citenamefont {Xing}\ \emph {et~al.}(2017)\citenamefont {Xing},
  \citenamefont {Zhou}, \citenamefont {Wang}, \citenamefont {Zhu},
  \citenamefont {Zhang}, \citenamefont {Zhou}, \citenamefont {Qian},
  \citenamefont {Xu},\ and\ \citenamefont {Shi}}]{xing2017two}%
  \BibitemOpen
  \bibfield  {author} {\bibinfo {author} {\bibfnamefont {X.}~\bibnamefont
  {Xing}}, \bibinfo {author} {\bibfnamefont {W.}~\bibnamefont {Zhou}}, \bibinfo
  {author} {\bibfnamefont {J.}~\bibnamefont {Wang}}, \bibinfo {author}
  {\bibfnamefont {Z.}~\bibnamefont {Zhu}}, \bibinfo {author} {\bibfnamefont
  {Y.}~\bibnamefont {Zhang}}, \bibinfo {author} {\bibfnamefont
  {N.}~\bibnamefont {Zhou}}, \bibinfo {author} {\bibfnamefont {B.}~\bibnamefont
  {Qian}}, \bibinfo {author} {\bibfnamefont {X.}~\bibnamefont {Xu}}, \ and\
  \bibinfo {author} {\bibfnamefont {Z.}~\bibnamefont {Shi}},\ }\href@noop {}
  {\bibfield  {journal} {\bibinfo  {journal} {Sci. Rep.}\ }\textbf {\bibinfo
  {volume} {7}},\ \bibinfo {pages} {45943} (\bibinfo {year}
  {2017})}\BibitemShut {NoStop}%
\bibitem [{\citenamefont {Hardy}\ \emph {et~al.}(2020)\citenamefont {Hardy},
  \citenamefont {Doussoulin}, \citenamefont {Klein}, \citenamefont {He},
  \citenamefont {Demuer}, \citenamefont {Willa}, \citenamefont {Willa},
  \citenamefont {Haghighirad}, \citenamefont {Wolf}, \citenamefont {Merz} \emph
  {et~al.}}]{hardy2020vortex}%
  \BibitemOpen
  \bibfield  {author} {\bibinfo {author} {\bibfnamefont {F.}~\bibnamefont
  {Hardy}}, \bibinfo {author} {\bibfnamefont {L.}~\bibnamefont {Doussoulin}},
  \bibinfo {author} {\bibfnamefont {T.}~\bibnamefont {Klein}}, \bibinfo
  {author} {\bibfnamefont {M.}~\bibnamefont {He}}, \bibinfo {author}
  {\bibfnamefont {A.}~\bibnamefont {Demuer}}, \bibinfo {author} {\bibfnamefont
  {R.}~\bibnamefont {Willa}}, \bibinfo {author} {\bibfnamefont
  {K.}~\bibnamefont {Willa}}, \bibinfo {author} {\bibfnamefont {A.-A.}\
  \bibnamefont {Haghighirad}}, \bibinfo {author} {\bibfnamefont
  {T.}~\bibnamefont {Wolf}}, \bibinfo {author} {\bibfnamefont {M.}~\bibnamefont
  {Merz}},  \emph {et~al.},\ }\href@noop {} {\bibfield  {journal} {\bibinfo
  {journal} {Phys. Rev. Research}\ }\textbf {\bibinfo {volume} {2}},\ \bibinfo
  {pages} {033319} (\bibinfo {year} {2020})}\BibitemShut {NoStop}%
\bibitem [{\citenamefont {Cui}\ and\ \citenamefont
  {Yang}(2008)}]{cui2008fulde}%
  \BibitemOpen
  \bibfield  {author} {\bibinfo {author} {\bibfnamefont {Q.}~\bibnamefont
  {Cui}}\ and\ \bibinfo {author} {\bibfnamefont {K.}~\bibnamefont {Yang}},\
  }\href@noop {} {\bibfield  {journal} {\bibinfo  {journal} {Phys. Rev. B}\
  }\textbf {\bibinfo {volume} {78}},\ \bibinfo {pages} {054501} (\bibinfo
  {year} {2008})}\BibitemShut {NoStop}%
\bibitem [{\citenamefont {Vorontsov}\ \emph {et~al.}(2005)\citenamefont
  {Vorontsov}, \citenamefont {Sauls},\ and\ \citenamefont
  {Graf}}]{vorontsov2005phase}%
  \BibitemOpen
  \bibfield  {author} {\bibinfo {author} {\bibfnamefont {A.}~\bibnamefont
  {Vorontsov}}, \bibinfo {author} {\bibfnamefont {J.}~\bibnamefont {Sauls}}, \
  and\ \bibinfo {author} {\bibfnamefont {M.}~\bibnamefont {Graf}},\ }\href@noop
  {} {\bibfield  {journal} {\bibinfo  {journal} {Phys. Rev. B}\ }\textbf
  {\bibinfo {volume} {72}},\ \bibinfo {pages} {184501} (\bibinfo {year}
  {2005})}\BibitemShut {NoStop}%
\bibitem [{\citenamefont {Ptok}(2010)}]{ptok2010fulde}%
  \BibitemOpen
  \bibfield  {author} {\bibinfo {author} {\bibfnamefont {A.}~\bibnamefont
  {Ptok}},\ }\href@noop {} {\bibfield  {journal} {\bibinfo  {journal} {Acta
  Phys. Pol. A}\ }\textbf {\bibinfo {volume} {118}},\ \bibinfo {pages} {420}
  (\bibinfo {year} {2010})}\BibitemShut {NoStop}%
\bibitem [{\citenamefont {Agterberg}\ and\ \citenamefont
  {Yang}(2001)}]{agterberg2001effect}%
  \BibitemOpen
  \bibfield  {author} {\bibinfo {author} {\bibfnamefont {D.}~\bibnamefont
  {Agterberg}}\ and\ \bibinfo {author} {\bibfnamefont {K.}~\bibnamefont
  {Yang}},\ }\href@noop {} {\bibfield  {journal} {\bibinfo  {journal} {J.
  Phys.: Condens. Matter}\ }\textbf {\bibinfo {volume} {13}},\ \bibinfo {pages}
  {9259} (\bibinfo {year} {2001})}\BibitemShut {NoStop}%
\bibitem [{\citenamefont {Wang}\ \emph {et~al.}(2007)\citenamefont {Wang},
  \citenamefont {Hu},\ and\ \citenamefont {Ting}}]{wang2007impurity}%
  \BibitemOpen
  \bibfield  {author} {\bibinfo {author} {\bibfnamefont {Q.}~\bibnamefont
  {Wang}}, \bibinfo {author} {\bibfnamefont {C.-R.}\ \bibnamefont {Hu}}, \ and\
  \bibinfo {author} {\bibfnamefont {C.-S.}\ \bibnamefont {Ting}},\ }\href@noop
  {} {\bibfield  {journal} {\bibinfo  {journal} {Phys. Rev. B}\ }\textbf
  {\bibinfo {volume} {75}},\ \bibinfo {pages} {184515} (\bibinfo {year}
  {2007})}\BibitemShut {NoStop}%
\bibitem [{\citenamefont {Ptok}(2014)}]{ptok2014influence}%
  \BibitemOpen
  \bibfield  {author} {\bibinfo {author} {\bibfnamefont {A.}~\bibnamefont
  {Ptok}},\ }\href@noop {} {\bibfield  {journal} {\bibinfo  {journal} {Eur.
  Phys. J. B}\ }\textbf {\bibinfo {volume} {87}},\ \bibinfo {pages} {2}
  (\bibinfo {year} {2014})}\BibitemShut {NoStop}%
\bibitem [{\citenamefont {Kenzelmann}(2017)}]{kenzelmann2017exotic}%
  \BibitemOpen
  \bibfield  {author} {\bibinfo {author} {\bibfnamefont {M.}~\bibnamefont
  {Kenzelmann}},\ }\href@noop {} {\bibfield  {journal} {\bibinfo  {journal}
  {Rep. Prog. Phys.}\ }\textbf {\bibinfo {volume} {80}},\ \bibinfo {pages}
  {034501} (\bibinfo {year} {2017})}\BibitemShut {NoStop}%
\bibitem [{\citenamefont {Suzuki}\ \emph {et~al.}(2011)\citenamefont {Suzuki},
  \citenamefont {Ichioka},\ and\ \citenamefont {Machida}}]{suzuki2011theory}%
  \BibitemOpen
  \bibfield  {author} {\bibinfo {author} {\bibfnamefont {K.~M.}\ \bibnamefont
  {Suzuki}}, \bibinfo {author} {\bibfnamefont {M.}~\bibnamefont {Ichioka}}, \
  and\ \bibinfo {author} {\bibfnamefont {K.}~\bibnamefont {Machida}},\
  }\href@noop {} {\bibfield  {journal} {\bibinfo  {journal} {Phys. Rev. B}\
  }\textbf {\bibinfo {volume} {83}},\ \bibinfo {pages} {140503} (\bibinfo
  {year} {2011})}\BibitemShut {NoStop}%
\bibitem [{\citenamefont {Ikeda}\ \emph {et~al.}(2010)\citenamefont {Ikeda},
  \citenamefont {Hatakeyama},\ and\ \citenamefont
  {Aoyama}}]{ikeda2010antiferromagnetic}%
  \BibitemOpen
  \bibfield  {author} {\bibinfo {author} {\bibfnamefont {R.}~\bibnamefont
  {Ikeda}}, \bibinfo {author} {\bibfnamefont {Y.}~\bibnamefont {Hatakeyama}}, \
  and\ \bibinfo {author} {\bibfnamefont {K.}~\bibnamefont {Aoyama}},\
  }\href@noop {} {\bibfield  {journal} {\bibinfo  {journal} {Phys. Rev. B}\
  }\textbf {\bibinfo {volume} {82}},\ \bibinfo {pages} {060510} (\bibinfo
  {year} {2010})}\BibitemShut {NoStop}%
\bibitem [{\citenamefont {Sun}\ \emph {et~al.}(2017)\citenamefont {Sun},
  \citenamefont {Kittaka}, \citenamefont {Nakamura}, \citenamefont
  {Sakakibara}, \citenamefont {Irie}, \citenamefont {Nomoto}, \citenamefont
  {Machida}, \citenamefont {Chen},\ and\ \citenamefont {Tamegai}}]{sun2017gap}%
  \BibitemOpen
  \bibfield  {author} {\bibinfo {author} {\bibfnamefont {Y.}~\bibnamefont
  {Sun}}, \bibinfo {author} {\bibfnamefont {S.}~\bibnamefont {Kittaka}},
  \bibinfo {author} {\bibfnamefont {S.}~\bibnamefont {Nakamura}}, \bibinfo
  {author} {\bibfnamefont {T.}~\bibnamefont {Sakakibara}}, \bibinfo {author}
  {\bibfnamefont {K.}~\bibnamefont {Irie}}, \bibinfo {author} {\bibfnamefont
  {T.}~\bibnamefont {Nomoto}}, \bibinfo {author} {\bibfnamefont
  {K.}~\bibnamefont {Machida}}, \bibinfo {author} {\bibfnamefont
  {J.}~\bibnamefont {Chen}}, \ and\ \bibinfo {author} {\bibfnamefont
  {T.}~\bibnamefont {Tamegai}},\ }\href@noop {} {\bibfield  {journal} {\bibinfo
   {journal} {Phys. Rev. B}\ }\textbf {\bibinfo {volume} {96}},\ \bibinfo
  {pages} {220505} (\bibinfo {year} {2017})}\BibitemShut {NoStop}%
\bibitem [{\citenamefont {Sun}\ \emph {et~al.}(2018)\citenamefont {Sun},
  \citenamefont {Kittaka}, \citenamefont {Nakamura}, \citenamefont
  {Sakakibara}, \citenamefont {Zhang}, \citenamefont {Shin}, \citenamefont
  {Irie}, \citenamefont {Nomoto}, \citenamefont {Machida}, \citenamefont
  {Chen},\ and\ \citenamefont {Tamegai}}]{sun2018disorder}%
  \BibitemOpen
  \bibfield  {author} {\bibinfo {author} {\bibfnamefont {Y.}~\bibnamefont
  {Sun}}, \bibinfo {author} {\bibfnamefont {S.}~\bibnamefont {Kittaka}},
  \bibinfo {author} {\bibfnamefont {S.}~\bibnamefont {Nakamura}}, \bibinfo
  {author} {\bibfnamefont {T.}~\bibnamefont {Sakakibara}}, \bibinfo {author}
  {\bibfnamefont {P.}~\bibnamefont {Zhang}}, \bibinfo {author} {\bibfnamefont
  {S.}~\bibnamefont {Shin}}, \bibinfo {author} {\bibfnamefont {K.}~\bibnamefont
  {Irie}}, \bibinfo {author} {\bibfnamefont {T.}~\bibnamefont {Nomoto}},
  \bibinfo {author} {\bibfnamefont {K.}~\bibnamefont {Machida}}, \bibinfo
  {author} {\bibfnamefont {J.}~\bibnamefont {Chen}}, \ and\ \bibinfo {author}
  {\bibfnamefont {T.}~\bibnamefont {Tamegai}},\ }\href@noop {} {\bibfield
  {journal} {\bibinfo  {journal} {Phys. Rev. B}\ }\textbf {\bibinfo {volume}
  {98}},\ \bibinfo {pages} {064505} (\bibinfo {year} {2018})}\BibitemShut
  {NoStop}%
\bibitem [{\citenamefont {Hanaguri}\ \emph {et~al.}(2018)\citenamefont
  {Hanaguri}, \citenamefont {Iwaya}, \citenamefont {Kohsaka}, \citenamefont
  {Machida}, \citenamefont {Watashige}, \citenamefont {Kasahara}, \citenamefont
  {Shibauchi},\ and\ \citenamefont {Matsuda}}]{hanaguri2018two}%
  \BibitemOpen
  \bibfield  {author} {\bibinfo {author} {\bibfnamefont {T.}~\bibnamefont
  {Hanaguri}}, \bibinfo {author} {\bibfnamefont {K.}~\bibnamefont {Iwaya}},
  \bibinfo {author} {\bibfnamefont {Y.}~\bibnamefont {Kohsaka}}, \bibinfo
  {author} {\bibfnamefont {T.}~\bibnamefont {Machida}}, \bibinfo {author}
  {\bibfnamefont {T.}~\bibnamefont {Watashige}}, \bibinfo {author}
  {\bibfnamefont {S.}~\bibnamefont {Kasahara}}, \bibinfo {author}
  {\bibfnamefont {T.}~\bibnamefont {Shibauchi}}, \ and\ \bibinfo {author}
  {\bibfnamefont {Y.}~\bibnamefont {Matsuda}},\ }\href@noop {} {\bibfield
  {journal} {\bibinfo  {journal} {Sci. Adv.}\ }\textbf {\bibinfo {volume}
  {4}},\ \bibinfo {pages} {eaar6419} (\bibinfo {year} {2018})}\BibitemShut
  {NoStop}%
\end{thebibliography}%

\pagebreak
\newpage
\onecolumngrid
\begin{center}
	\textbf{\huge Supplemental information}
\end{center}
\vspace{1cm}
\onecolumngrid
\setcounter{equation}{0}
\setcounter{figure}{0}
\setcounter{table}{0}

\makeatletter
\renewcommand{\theequation}{S\arabic{equation}}
\renewcommand{\thefigure}{S\arabic{figure}}

\subsection*{S1 \textit{$B$}$_{\rm{c}2}$ determined by the criterion of the onset of the SC transition}

Figure S3 shows the temperature-dependent \textit{$B$}$_{\rm{c}2}$ determined by the criterion of the onset of the superconducting (SC) transition. Symbols of the diamond and the circle represent the case of \textit{B} $\parallel$ \textit{c} and \textit{B} $\perp$ \textit{c}, respectively. The open symbols are obtained by scanning field (\textit{B} scan), and the closed symbols are obtained by scanning temperature (\textit{T} scan). The two-band model (green lines) and the WHH fitting curves (black lines) are shown for \textit{B} $\parallel$ \textit{c} and \textit{B} $\perp$ \textit{c}. Meanwhile, the WHH model predictions with \textit{$\alpha$} = 0 are also presented for comparison. We observe the upper critical field \textit{B}$_{\rm{c}2}^{\rm{onset}}$(\textit{T}) data significantly deviates from the WHH curves, and displays an unusual upturn for all the three crystals as marked by the pink color.

\subsection*{S2 The estimation of the coherence length}

Details about the estimation of coherence length are shown in Fig. S4 and Fig. S5. Fig. S4 shows the temperature-dependent upper critical field \textit{B}$_{\rm{c}2,\textit{ab}}^{50\%}$(\textit{T}) for the three samples. The open symbols are obtained by scanning field (\textit{B} scan), and the closed symbols are obtained by scanning temperature (\textit{T} scan). The zero-temperature coherence length was estimated by using \textit{$\xi$}$_{\textit{ab}}$(0) = ($\Phi$$_0$/2\textit{$\pi$}\textit{B}$_{\rm{c}2}^{\textit{c}}$(0))$^{1/2}$, and \textit{$\xi$}$_{\textit{c}}$(0) = $\Phi$$_0$/2\textit{$\pi$}\textit{$\xi$}$_{\textit{ab}}$(0)\textit{B}$_{\rm{c}2}^{\textit{ab}}$(0) ($\Phi$$_0$ = 2.07$\times$10$^{-15}$ Wb is the magnetic flux quantum). Under low temperature region, the \textit{B}$_{\rm{c}2}^{\parallel {\textit{ab}}}$(0 K) is determined by linearly extrapolating the upper critical field data to zero temperature, which obtains the \textit{$\xi$}$_{\textit{c}}$(0) $\sim$ 2.6 nm (S1), 2.7 nm (S2), and 2.9 nm (S3), respectively. Meanwhile, we also extract the \textit{B}$_{\rm{c}2}^{\textit{c}}$(0) from the two-band fitting for estimating the coherence lengths, with \textit{$\xi$}$_{\textit{ab}}$(0) $\sim$ 4.7 nm (S1), 4.8 nm (S2), and 5.1 nm (S3), respectively. Furthermore, the coherence lengths are also estimated by using the anisotropic Ginzburg-Landau equations (see Fig. S5). The coherence lengths is determined by using the anisotropic Ginzburg-Landau (AGL) expressions \textit{$\xi$}$_{\textit{ab}}$ = ($\Phi$$_{0}$/2\textit{$\pi$}\textit{B}$_{\rm{c}2}^{\textit{c}}$)$^{1/2}$ and \textit{$\xi$}$_{\textit{c}}$ = $\Phi$$_{0}$/2\textit{$\pi$}\textit{$\xi$}$_{\textit{ab}}$\textit{B}$_{\rm{c}2}^{\textit{ab}}$. The solid lines are the fits of the AGL equations \textit{$\xi$}$_{\textit{ab}}$(\textit{T}) = \textit{$\xi$}$_{\textit{ab}}$(0) $\times$ (1-\textit{T}/\textit{T}$_{\rm{c}}$)$^{-1/2}$ and \textit{$\xi$}$_{\textit{c}}$(\textit{T}) = \textit{$\xi$}$_{\textit{c}}$(0) $\times$ (1-\textit{T}/\textit{T}$_{\rm{c}}$)$^{-1/2}$. We obtain the \textit{$\xi$}$_{\textit{ab}}$(0) $\sim$ 4.45 nm (S1), 4.52 nm (S2), and 4.8 nm (S3), which is in reasonable agreement with the results derived from the two-band model. However, for \textit{$\xi$}$_{\textit{c}}$, the AGL fitting shows an obvious deviation at low temperatures.

\subsection*{S3 Temperature dependent \textit{$B$}$_{\rm{c}2}$ determined by the resistivity (end point) and torque magnetometry}

In order to confirm the behaviour of the upper critical field \textit{B}$_{\rm{c}2}$, we also performed the measurements of magnetic torque under static fields up to 30 T on the same piece of crystals for resistivity measurements (see Figs. S6(a)-S6(c)). The magnetic torque is a thermal dynamic measurements probing the bulk information of sample [37, 38].
From the torque data, we can obtain the irreversibility field \textit{$B$}$_{\rm{irr}}$ from the onset of the separating fields for the up- and down-sweeps torque data (as indicated by arrows in Figs. S6(a)-S6(c)). More details about the determination of \textit{$B$}$_{\rm{irr}}$ can be seen in the enlarged plot of the separating fields shown in Fig. S7. The extracted irreversibility field \textit{$B$}$_{\rm{irr}}$ is in good agreement with the \textit{$B$}$_{\rm{c}2}$(T) determined by the criteria of the end point of the superconducting transition as expected (see Figs. S8(a)-S8(c)). This result confirms that the obtained \textit{$B$}$_{\rm{c}2}$ is reliable. \textit{$B$}$_{\rm{c}2}^{\rm{endpoint}}$ normalized by the \textit{$B$}$_{\rm{c}2,\textit{ab}}^{\rm{WHH}}$(0 K) (WHH fitting with finite \textit{$\alpha$}) is shown in Fig. S8(d). For comparison, \textit{$B$}$_{\rm{c}2}^{\textit{ab}}$ obtained by the resistivity reported by T. Terashima \textit{et al}. (open squares) [36], S. Kasahara \textit{et al}. (open circles) [20], J. M. Ok \textit{et al}. (open diamonds) [19], obtained by the thermal conductivity reported by S. Kasahara \textit{et al}. (open triangles) [20], and obtained by the heat capacity reported by F. Hardy (open pentagons) [41] are also included. Clearly, all the data can be roughly scaled, although the scaling is slightly worse than those determined by 50\% of the SC transition (Fig. 3(d)), which is due to the influence of vortex motion.

\begin{figure}\center
\includegraphics[width=8.8cm]{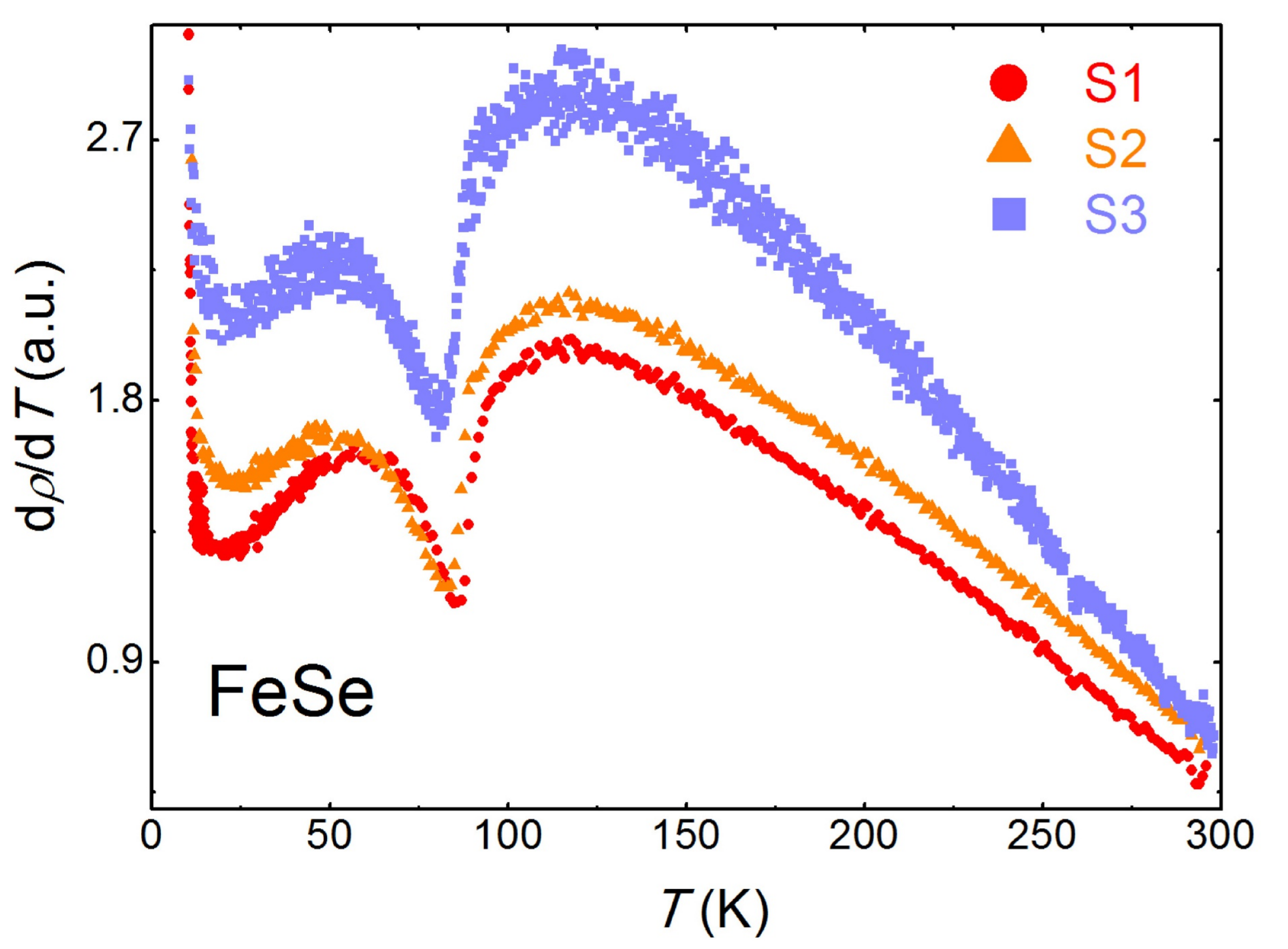}\\
\caption{\label{S1} Temperature dependencies of the first derivative of \textit{$\rho$}--\textit{T} for the three selected FeSe single crystals with different amounts of disorders.}\label{}
\end{figure}

\begin{figure}\center
\includegraphics[width=13.2cm]{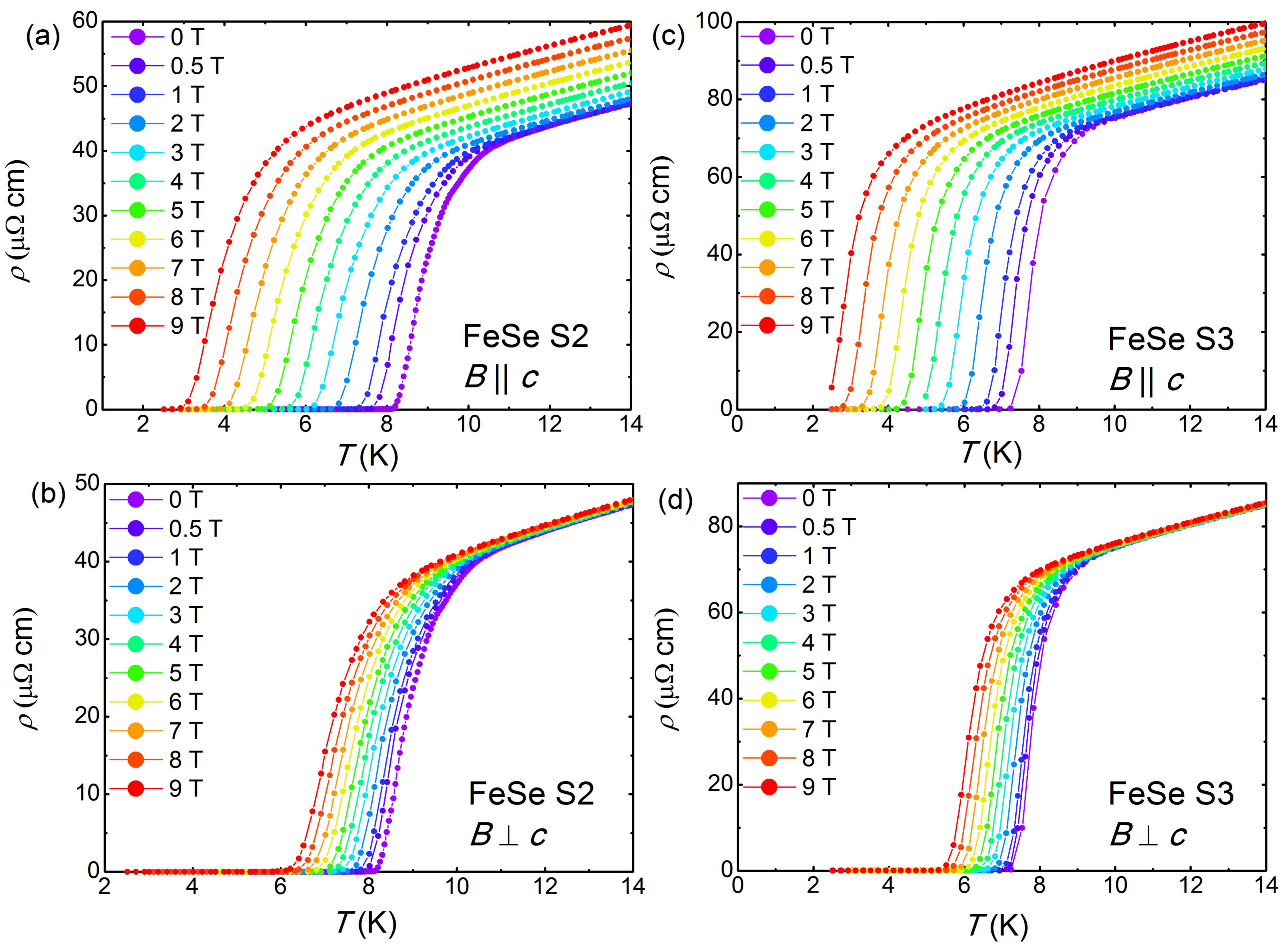}\\
\caption{\label{S2} Temperature dependencies of the resistivity with \textit{B} $\parallel$ \textit{c} and \textit{B} $\perp$ \textit{c} under fields from 0 to 9 T for (a-b) sample S2 and (c-d) sample S3.}\label{}
\end{figure}

\begin{figure}\center
\includegraphics[width=16cm]{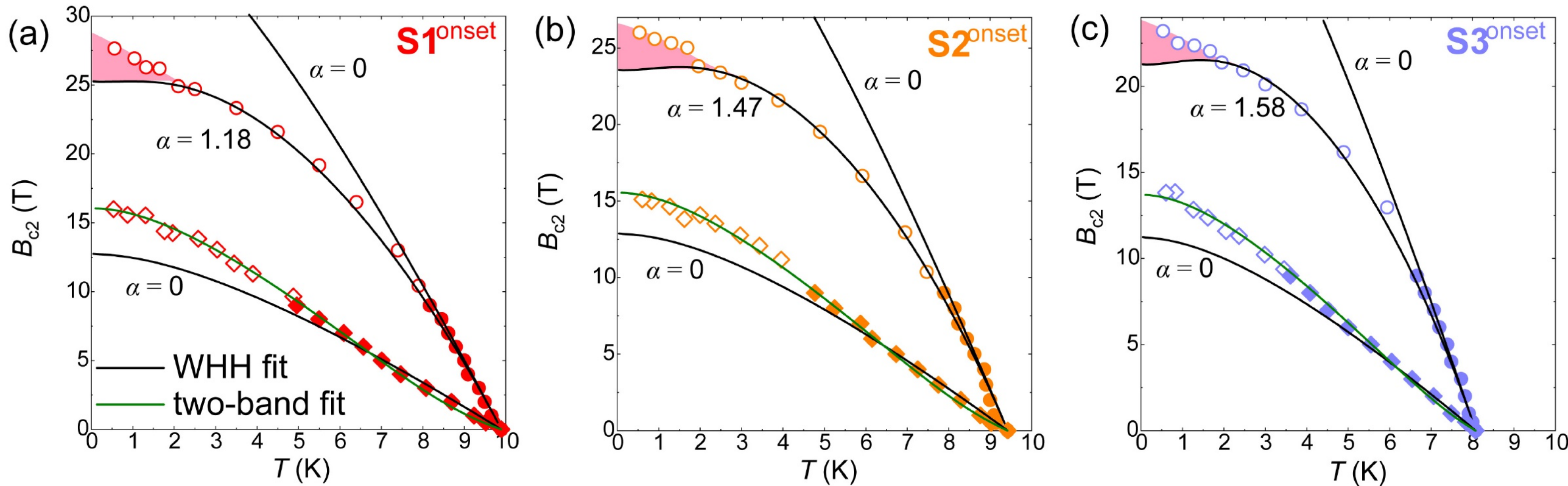}\\
\caption{\label{S3} Temperature dependencies of the upper critical field \textit{B}$_{\rm{c}2}^{\rm{onset}}$(\textit{T}) for the three selected FeSe single crystals.}\label{}
\end{figure}

\begin{figure}\center
\includegraphics[width=22pc]{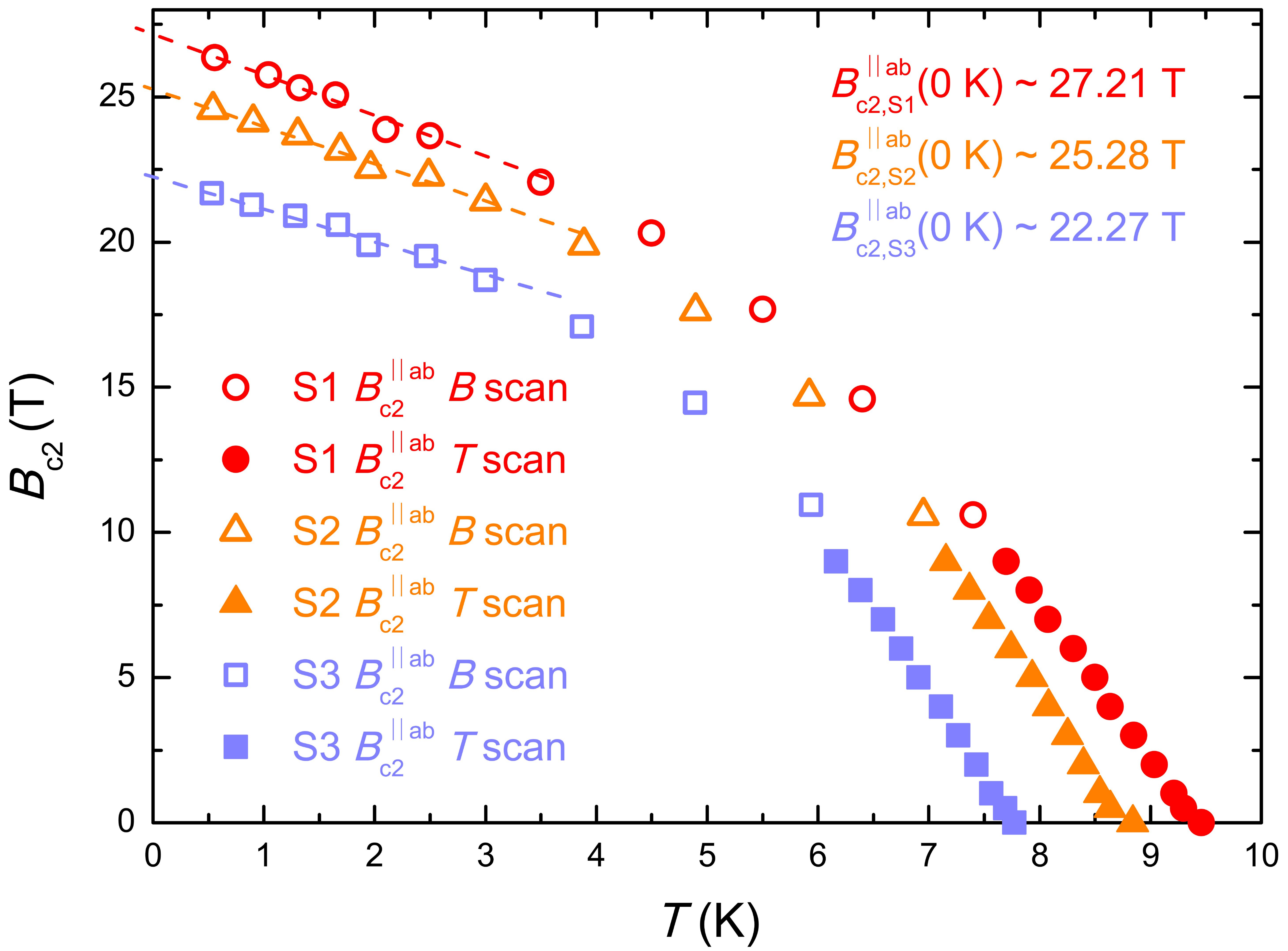}\\
\caption{\label{S4} Temperature dependencies of the upper critical field \textit{B}$_{\rm{c}2,\textit{ab}}^{50\%}$(\textit{T}) for the three selected FeSe single crystals with different amounts of disorders.}\label{}
\end{figure}

\begin{figure}\center
\includegraphics[width=25pc]{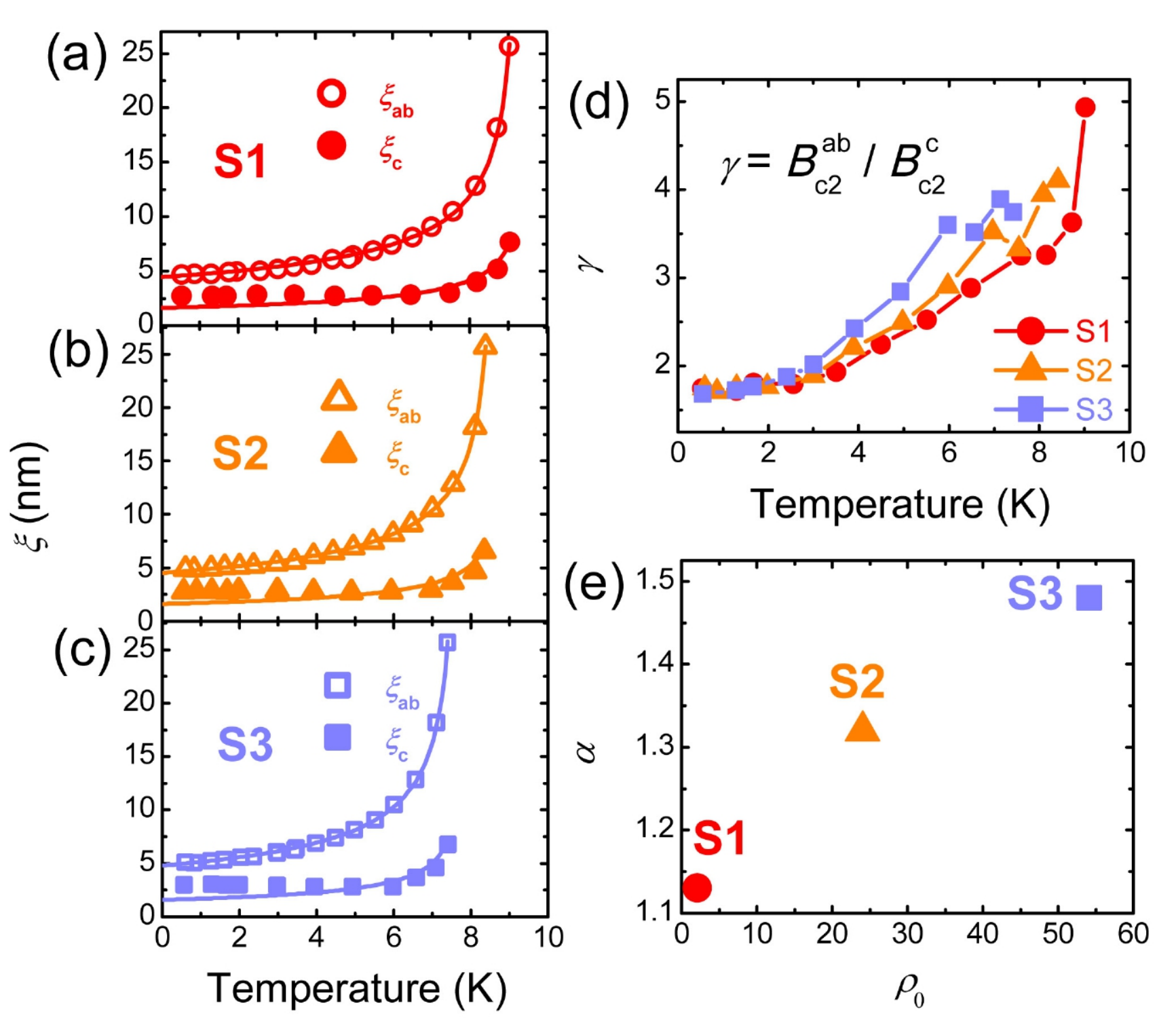}\\
\caption{\label{S5} [(a)--(c)] Temperature dependencies of the in-plane coherence length \textit{$\xi$}$_{\textit{ab}}$(\textit{T}) (open symbols) and out-of-plane coherence length \textit{$\xi$}$_{\textit{c}}$(\textit{T}) (solid symbols). (d) Temperature dependence of the anisotropy parameter \textit{$\gamma$} = \textit{B}$_{\rm{c}2}^{\textit{ab}}$/\textit{B}$_{\rm{c}2}^{\textit{c}}$. (e) The residual resistivity \textit{$\rho$}$_{0}$ dependence of Maki parameter \textit{$\alpha$}.}\label{}
\end{figure}

\begin{figure}\center
\includegraphics[width=38pc]{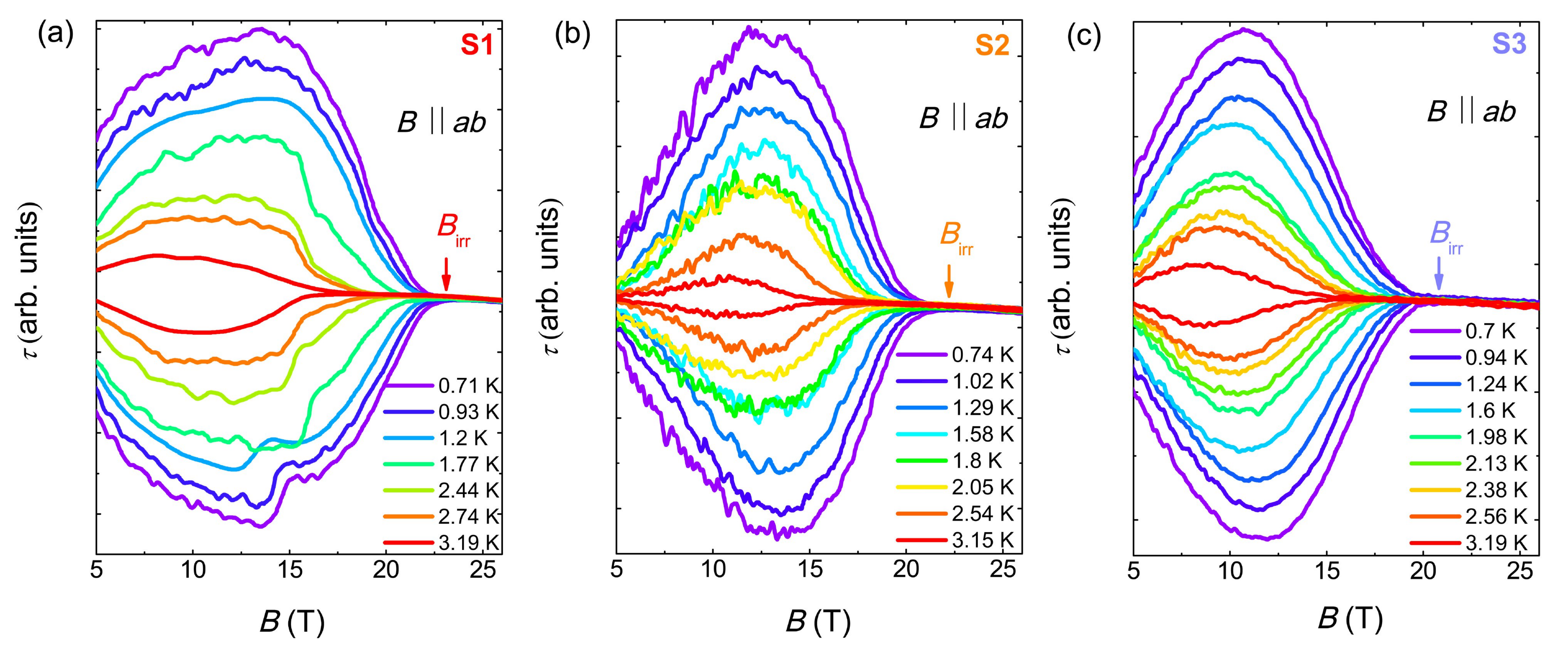}\\
\caption{\label{S6} Magnetic field dependencies of the magnetic torque \textit{$\tau$}(\textit{B}) with \textit{B} $\parallel$ \textit{ab} at various temperatures for (a) sample S1, (b) sample S2, (c) sample S3.}\label{}
\end{figure}

\begin{figure}\center
\includegraphics[width=42pc]{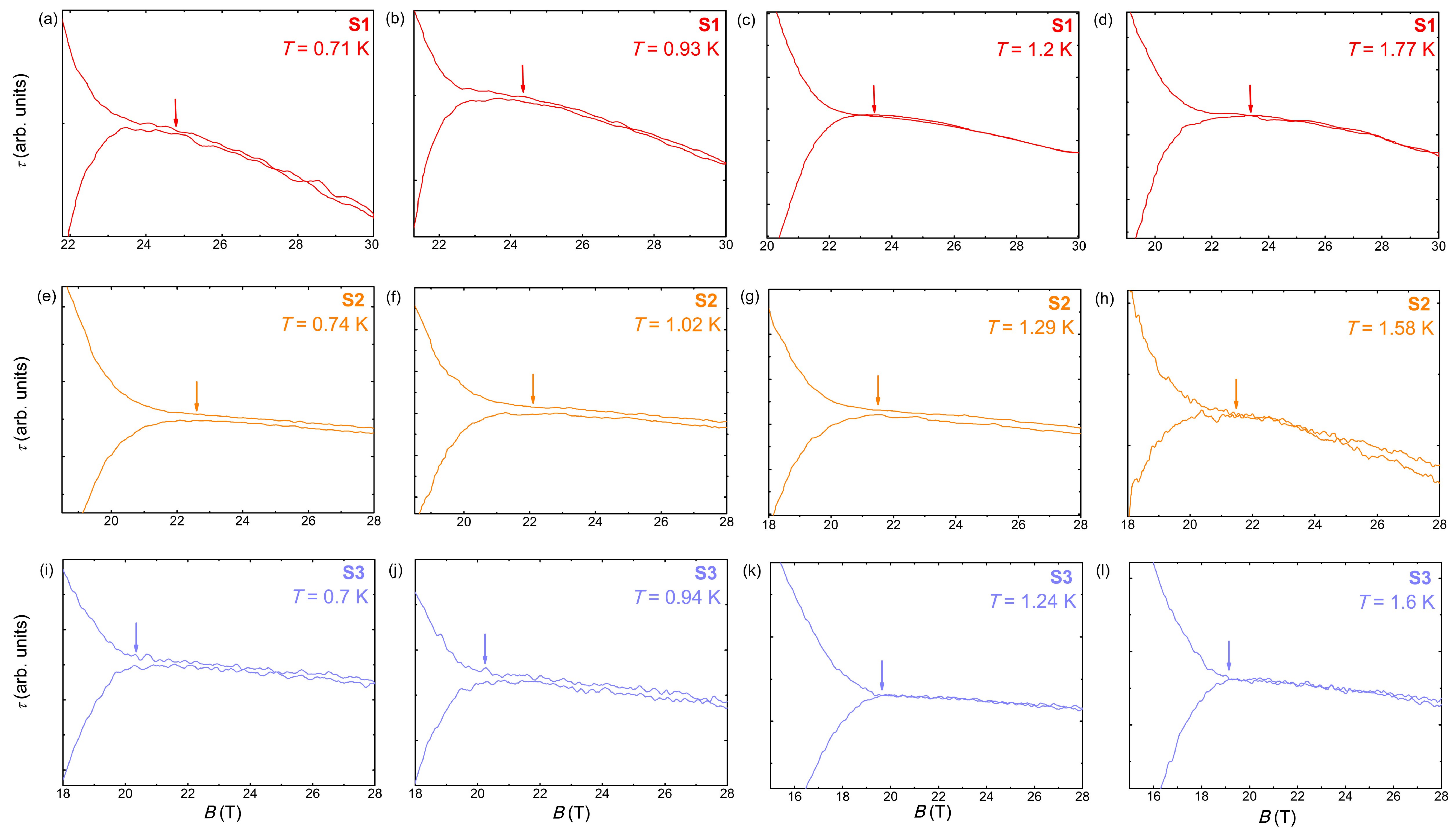}\\
\caption{\label{S7} The determinations of the irreversibility field \textit{B}$_{\rm{irr}}$ for (a-d) sample S1, (e-h) sample S2, (i-l) sample S3, which were  defined as the onset of the separating field for the up- and down-sweeps torque data.}\label{}
\end{figure}

\begin{figure}\center
\includegraphics[width=20pc]{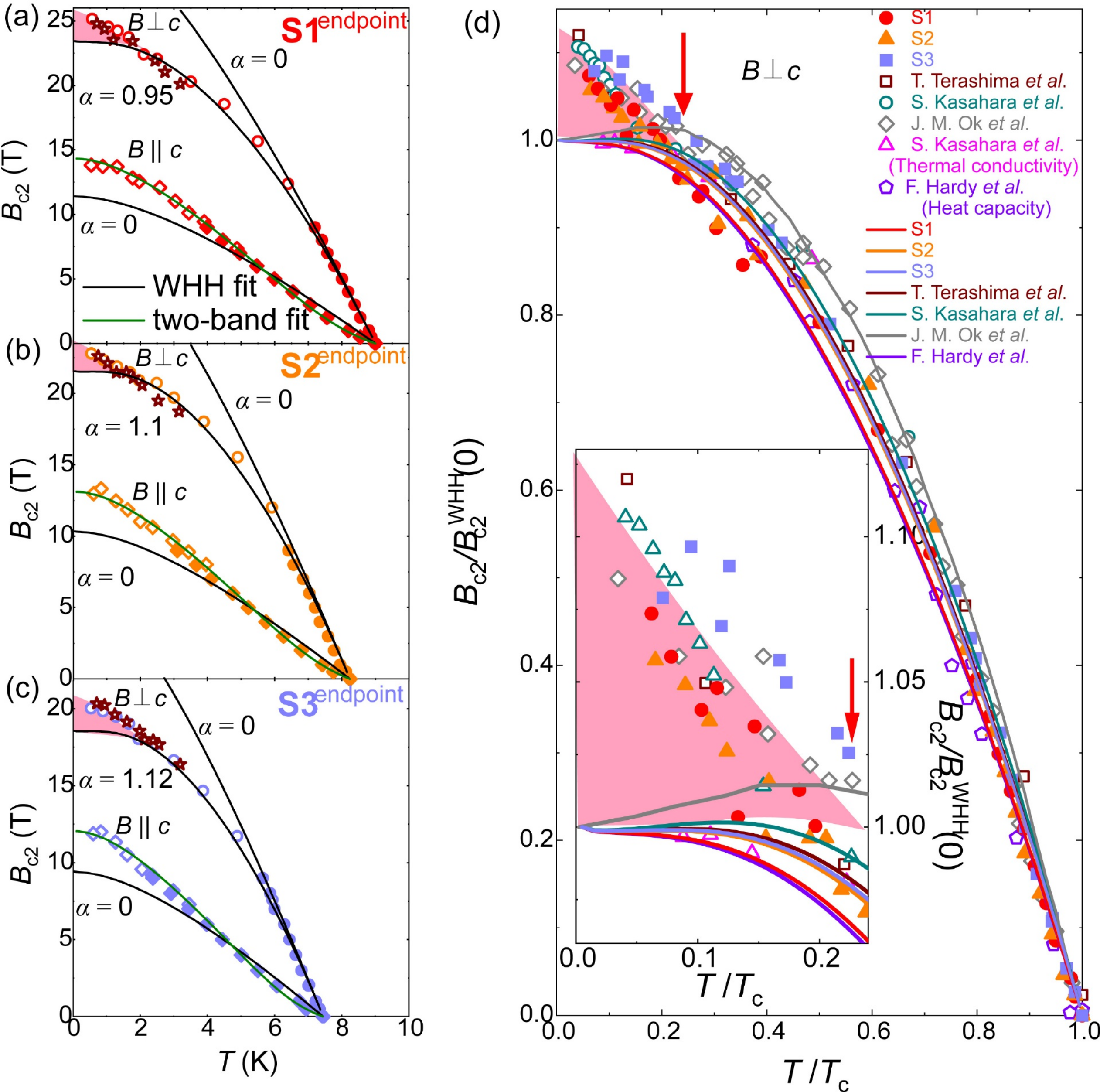}\\
\caption{\label{S8} [(a)--(c)] Temperature dependencies of the upper critical field \textit{B}$_{\rm{c}2}$(\textit{T}) for the three selected FeSe single crystals. \textit{B}$_{\rm{c}2}$(\textit{T}) are determined by resistivity (diamond and circle) and torque magnetometry (open star) measurements. The resistivity measurements correspond to the criteria of the end point of the superconducting transition. Symbols of the diamond and circle represent the case of \textit{B} $\parallel$ \textit{c} and \textit{B} $\perp$ \textit{c}, respectively. The two-band model (green lines) and WHH fitting curves (black lines) are shown for \textit{B} $\parallel$ \textit{c} and \textit{B} $\perp$ \textit{c}. Meanwhile, the WHH model predictions with \textit{$\alpha$} = 0 are also presented for comparison. (d) The critical field \textit{B}$_{\rm{c}2}$ normalized by the \textit{$B$}$_{\rm{c}2}^{\rm{WHH}}$(0 K) for the crystals used in this paper and these previous works [19, 20, 36, 41]. The solid lines represent the WHH fitting. The inset shows the enlarged low temperature part.}\label{}
\end{figure}

\end{document}